\documentclass[prb,twocolumn,showpacs,floatfix,amsmath,eqsecnum]{revtex4}
\usepackage{epsfig}

\begin{document}

\title{Conductance anomalies and the 
extended Anderson model\\ for nearly perfect quantum wires}

\author{T. Rejec$^1$ and A. Ram\v sak$^{1,2}$}
\affiliation{$^1$J. Stefan Institute, 1001 Ljubljana, Slovenia\\
$^{2}$ Faculty of Mathematics and Physics, University of
Ljubljana, 1001 Ljubljana, Slovenia }
\author{J.H. Jefferson}
\affiliation{QinetiQ, Sensors and Electronic Division,
St. Andrews Road, Great Malvern,\\
Worcestershire WR14 3PS, England}

\date{\today}
            
\begin{abstract}
Anomalies near the conductance threshold of nearly perfect
semiconductor quantum wires are explained in terms of singlet and
triplet resonances of conduction electrons with a single weakly-bound
electron in the wire. This is shown to be a universal effect for a
wide range of situations in which the effective single-electron
confinement is weak. The robustness of this generic behavior is
investigated numerically for a wide range of shapes and sizes of
cylindrical wires with a bulge. The dependence on gate voltage,
source-drain voltage and magnetic field is discussed within the
framework of an extended Hubbard model. This model is mapped onto an
extended Anderson model, which in the limit of low temperatures is
expected to lead to Kondo resonance physics and pronounced many-body
effects.
\end{abstract}   


\maketitle
  

\section{Introduction}
Semiconductor quantum wires can be fabricated with effective wire
widths down to a few nanometers; for example, by heteroepitaxial
growth on `v'-groove surfaces \cite{walther92} and ridges
\cite{ramvall97}, cleaved edge over-growth \cite{yacoby96}, etched
wires with gating \cite{kristensen98}, and gated two-dimensional
electron gas (2DEG) structures \cite{wees88,wharam88}. More recently,
there has been considerable interest in carbon nanotubes for which the
quantum wire cross section can approach atomic dimensions. Such
structures have potential for opto-electronic applications, such as
light-emitting diodes, low-threshold lasers, single-electron devices
and quantum information processing.

Conductance steps in various types of quantum point contacts and
quantum wires were found more than a decade ago
\cite{wees88,wharam88}. These first experiments are broadly consistent
with a simple non-interacting picture\cite{houten92}. However, there
are certain anomalies, some of which are believed to be related to
electron-electron interactions and appear to be spin-dependent. In
particular, a structure is seen in the rising edge of the conductance
curve, starting at around $0.7(2e^{2}/h)$ and merging with the first
conductance plateau with increasing energy \cite{thomas96}. This
structure, already visible in the early experiments \cite{wees88}, can
survive to temperatures of a few degrees and also persists under
increasing source-drain bias, even when the conductance plateau has
disappeared. Under increasing in-plane magnetic field, the structure
moves down, eventually merging with the $e^{2}/h$ conductance plateau
at very high fields and is not a transmission effect through a
ballistic channel \cite{liang}.  A structure is seen also in high
quality quantum wires \cite{pyshkin}. In some experiments, an anomaly
is seen at lower energy with conductance around $0.3(2e^{2}/h)$
\cite{kaufman99,ramvall97}. This can also survive to a few degrees,
though is less robust than the 0.7 anomaly and is more readily
suppressed by a magnetic field \cite{ramvall97}. Recently the anomaly
was confirmed also in back-gated \cite{nuttinck}, in shallow-etched
\cite{kristensen00} point contacts and in a ballistic quantum
wire\cite{liang00}. At low temperatures the anomaly exhibits a
puzzling similarity with Kondo resonance behavior \cite{kondo2002},
as do thermopower measurements\cite{appleyard00}.

Theoretical work has attempted to explain these observations in various
ways, including conductance suppression in a Luttinger liquid with
repulsive interaction and disorder \cite{maslov95}, local
spin-polarized density-functional theory \cite{wang98} and
spin-polarized sub-bands \cite{fasol94}. Near the conduction
threshold, there is a 'Coulomb blockade' and we have shown that this
gives rise to spin-dependent resonances, for wires of both rectangular
\cite{rejec002d} and cylindrical \cite{rejec003d} cross-section, with
related anomalies in thermoelectric transport coefficients
\cite{rejec02}. A similar singlet-triplet scenario was presented in 
Ref.~\onlinecite{flambaum00} and a phenomenological approach is presented in
Ref.~\onlinecite{landau}. Recent studies have investigated the
0.7 anomaly in quantum point
contacts within the Hartree-Fock approximation \cite{sushkov1},
spin-fluctuation backscattering \cite{tokura} and in the framework of the 
Anderson model with related Kondo resonance behavior \cite{meir2002}.

In Refs.~\onlinecite{rejec002d,rejec003d,rejec02} we suggested that these 
anomalies are related to weakly bound
states and resonant bound states within the wire. These would arise, for
example, from a small fluctuation in thickness of the wire in some
region giving rise to a weak bulge. If this bulge is very weak then only a
single electron will be bound. We may thus regard this system as an `open'
quantum dot in which the bound electron inhibits the transport of conduction
electrons via the Coulomb interaction. Near the conduction threshold, there
will be a Coulomb blockade and we show below that this also gives rise to a
resonance, analogous to that which occurs in the single-electron transistor 
\cite{meirav91}. This is a generic effect arising from an electron bound in
some region of the wire and such binding may arise from a number of sources,
which we do not consider explicitly. For example, in addition to a weak
thickness fluctuation, a smooth variation in confining potential due to
remote gates, contacts and depletion regions could contribute to electron
confinement along the wire or gated 2DEG. A significant contribution to the 
single-electron confinement could also arise from its electronic polarization
of the lattice or image charge.

In this paper we extend our previous study of a particular geometry of
the quantum wire with a comprehensive analysis of a wide range of
shapes and sizes of wire in order to demonstrate the generic and wide
applicability of the phenomena. We study in particular the threshold
of the conductivity of nearly perfect wires for which a single
electron is bound. We express the conductance in terms of the
two-electron scattering matrix. In order to extend the exact
two-electron analysis into the true many-electron domain, we construct
an extended Anderson model and analyze the influence of the
corresponding momentum dependent coupling matrix elements.

The model is introduced in the next section and the special case of a
cylindrical GaAs wire is derived in Appendix A.  In Section III a
detailed analysis of the two-electron problem is presented in which
one electron is weakly bound in the wire, giving rise to
spin-dependent scattering of the other. Exact singlet and triplet
scattering states are computed near the conductance threshold.  In
Section IV we then show how the solutions of the scattering problem
may be used to determine conductance by an extension of the
Landauer-B\"{u}ttiker formula. This gives excellent agreement with a
number of experiments on different kinds of quantum wire. The effect
of finite magnetic field on the anomalies is presented and it is shown
how they are related to the spin-split steps in perfect quantum wires. In
the last Section we also examine the dependence of the anomalies on
asymmetry introduced by finite source-drain voltage and
summarize. Additional appendices are devoted to technical details
on the solution of the two-electron wave function in an external
potential, and the Hartree-Fock analysis.

\section{Basic model}

\begin{figure}[htb]
{\par\centering \resizebox*{0.9\columnwidth}{!}{\rotatebox{0}
{\includegraphics{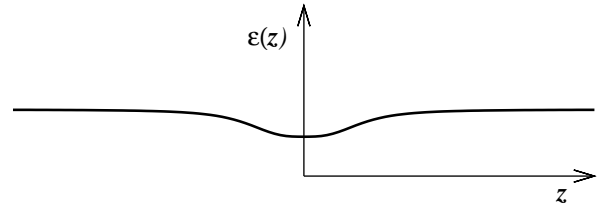}}} \par}
\caption{\label{land1} Effective one-dimensional well caused by
thickness fluctuation, impurity charge, gate image charge, self-polarization
due to single electron or some combination of these.}
\end{figure}

In previous work\cite{rejec002d,rejec003d,rejec02}, we have considered
a straight quantum wire with a small fluctuation in
thickness giving rise to a weak `bulge'. 
The precise details of the bulge are largely
unimportant for what follows, the main requirement being that the
change in the width of the wire is sufficiently gradual that
inter-channel mixing of the transverse modes is negligible and that
only one electron may be bound in the bulge region. The latter is
always the case for a weak symmetric bulge, which has at least one
bound state which can only sustain one electron due to Coulomb
repulsion. The problem reduces to electrons moving in an effective
weak potential well if we confine ourselves (by choice of gate
voltage) to the Fermi energies for which no more than one transverse
mode is occupied, i.e. the conductance threshold and the first
conductance step.  A typical effective potential well for such a bulge
is shown in
Fig.~\ref{land1}. Such a potential well may arise in other ways, such as
an actual potential fluctuation due to a nearby unscreened charged
impurity, or even some self-consistent effect due
to the electrons themselves through electronic polarization and image
charge in a remote gate.  We shall not consider
the possible cause of this weak potential further but emphasize that
because it may arise in many ways, the weak potential well model is
very general with widespread applicability.

Consider now the motion of electrons in the wire near the conductance
threshold. A single electron will be bound in the potential well
region and the remaining electrons will undergo scattering from the
localized electron via the Coulomb interaction as they propagate from
source to drain. At sufficiently low Fermi energy, the electrons in
the source contact will be totally reflected by the bound electron due
to Coulomb repulsion and there will be no current from source to drain
at $T=0$. As the Fermi energy is raised, the energy of the electrons
in the source contact will be sufficiently high for them to overcome
the Coulomb repulsion of the bound electron and a current will
flow. In calculating this current we will make the approximation that
the electrons flowing from source to drain only interact with the
bound electron via a screened Coulomb interaction. This is a
reasonable approximation provided that the electron density is not too
low in the region of interest, i.e. the rising edge to the first
conductance plateau. More precisely, the mean density of electrons in
the wire (number per unit length) should be at least of order the
inverse effective Bohr radius of the material. We return to this point
again in the final section.  Within this approximation, the
many-electron problem is reduced to an effective two-electron problem
in which one electron is bound and the other is a representative
electron at the Fermi energy in the leads. We show below that by
solving this two-electron problem exactly and summing over all
electrons near the Fermi energy we may compute the conductance.

\subsection{Extended Hubbard and Anderson model}

The Hamiltonian corresponding to interacting electrons in the wire
with a small geometric or potential inhomogeneity and close to the
threshold of conduction is, within the effective-mass
approximation an extended Hubbard Hamiltonian on a
finite-difference lattice \cite{rejecand},
\begin{equation}
\label{hamdis1d}
H=\sum _{\sigma }H_{1\sigma }+\frac{1}{2}\sum _{i\neq
j}U_{ij}n_{i}n_{j}+
\sum _{i}U_{ii}n_{i\uparrow }n_{i\downarrow }.
\end{equation}
Here  \( H_{1\sigma } \) is the single-particle 
Hamiltonian:
\begin{equation}
\label{ham1dis1d}
H_{1\sigma }=-t\sum _{i}\left( c_{i+1\sigma }^{\dagger }c_{i\sigma
}+c_{i\sigma }^{\dagger }
c_{i+1\sigma }\right) +\sum _{i}\epsilon _{i}n_{i\sigma ,}
\end{equation}
where \( c_{i\sigma }^{\dagger }, c_{i\sigma } \) are electron creation, 
annihilation operators,
\( n_{i\sigma }=c_{i\sigma }^{\dagger }c_{i\sigma } \), and
\( n_{i}=\sum _{\sigma }n_{i\sigma } \). Model parameters are 
hopping  $t$,  local potential at site \( i \),
$\epsilon _{i}$,
and screened electron-electron interaction at sites  \( i \) and \( j \),
$U_{ij}$. 
This Hamiltonian is derived and justified in Appendix A.

In order to study the many-electron problem, it is also convenient to
express the Hubbard Hamiltonian, Eq.~(\ref{hamdis1d}), in a basis
which distinguishes bound and unbound states
explicitly. Single-electron solutions corresponding to the
tight-binding Hamiltonian Eq.~(\ref{ham1dis1d}), follow from the
single-particle Schr\"odinger equation
\begin{equation}
H_{1}\left| \varphi \right\rangle =E_{1}\left| \varphi \right\rangle, \label{sin}
\end{equation}
and (with omitted spin index $\sigma$),
$\left| \varphi \right\rangle =\sum _{j}\varphi _{j}c_{j}^{\dagger
}\left| 0\right\rangle$. For large \( \left| j\right|  \) the potential 
\( \epsilon _{j} \) is
constant, therefore the solutions are asymptotically plane waves.
We thus diagonalize this
single-electron part of the Hamiltonian using the transformation
$c_{q\sigma }^\dagger=\sum_{j}c_{j\sigma }^\dagger \phi _{j}^{q}$,
where $\phi _{j}^{q}=\langle j| q \rangle \sim \exp ({i q j})$ asymptotically
for unbound states, with eigenenergies
$\varepsilon _{q}$. In this basis the Hamiltonian becomes,
\begin{equation}
H=\sum_{q}\epsilon _{q}n_{q}+\frac{1}{2}\!\!\!\!\!\sum\limits_{\scriptstyle q_1
q_2 q_3 q_4 \scriptstyle \sigma \sigma^\prime}\!\!\!\!{{\cal
U}(q_{1}q_{2}q_{3}q_{4})c_{q_{1}\sigma }^{\dagger }c_{q_{3}\sigma
^{\prime }}^{\dagger }c_{q_{4}\sigma ^{\prime }}c_{q_{2}\sigma }},
\label{Rejec_h2q}
\end{equation}
where 
\begin{equation}
{\cal U}(q_{1}q_{2}q_{3}q_{4})=\sum_{ij}U_{ij}(\phi
_{i}^{q_{1}})^{\ast }\phi _{i}^{q_{2}}(\phi _{j}^{q_{3}})^{\ast }\phi
_{j}^{q_{4}}.\label{uqq}
\end{equation}
We further denote the lowest bound state with
energy $\epsilon _{q}<0$ by $d_{\sigma }\equiv c_{q\sigma
},$ with $n_{d}=\sum_{\sigma }d_{\sigma }^{\dagger }d_{\sigma }$
and, similarly, the scattering states with positive $\epsilon _{q}$ are
distinguished by $q\rightarrow k$. There are two independent unbound states
corresponding to each $k$ and these are chosen to be plane waves
asymptotically, i.e. $\phi _{j}^{k}\rightarrow e^{ikj}$ as $j\rightarrow \pm
\infty $ and $\epsilon _{k}=\frac{\hbar ^{2}k^{2}}{2m^{\ast }}.$ Retaining
only those Coulomb matrix elements which involve both localized and
scattered electrons, omitting all terms which would give rise to states in
which the localized state is unoccupied, we arrive at an Anderson-type
Hamiltonian\cite{anderson61,rejecand},
\begin{eqnarray}
H &=&\sum_{k}\epsilon _{k}n_{k}+\epsilon _{d}n_{d}+ \sum_{k\sigma
}(V_{k}n_{d\bar{\sigma}}c_{k\sigma }^{\dagger }d_{\sigma }+
\mathrm{h.c.})+\\\nonumber
&+&Un_{d\uparrow }n_{d\downarrow } +\sum_{kk^{\prime
}\sigma }M_{kk^{\prime }}n_{d}c_{k\sigma }^{\dagger }c_{k^{\prime
}\sigma }+\sum_{kk^{\prime } }J_{kk^{\prime }}{\bf S }_{\bf d}\cdot
{\bf s}_{\bf kk^{\prime }}.
\label{Rejec_k} 
\end{eqnarray}
Here $U={\cal U}(dddd)$ is the Hubbard repulsion,
$V_{k}={\cal U} (dddk) $ is mixing term, $M_{kk^{\prime
}}={\cal U}(ddkk^{\prime })-\frac{
1}{2}{\cal U}(dkk^{\prime }d)$ corresponds to scattering of
electrons and the direct exchange coupling is $J_{kk^{\prime
}}=2\,{\cal U}(dkk^{\prime }d)$. Spin operators in
Eq.~(\ref{Rejec_k}) are defined as ${\bf S}_{\bf d}=\frac{1
}{2}\sum_{\sigma \sigma ^{\prime }}d_{\sigma }^{\dagger }{
\hbox{\boldmath$\sigma$}}_{\sigma \sigma ^{\prime }}d_{\sigma
^{\prime }}$ and ${\bf s}_{\bf kk^{\prime
}}=\frac{1}{2}\sum_{\sigma \sigma ^{\prime }}c_{k\sigma
}^{\dagger }{\hbox{\boldmath$\sigma$}}_{\sigma \sigma ^{\prime
}}c_{k^{\prime }\sigma ^{\prime }}$, where the components of
${\hbox{\boldmath$\sigma$}}$ are the
usual Pauli matrices. A similar model has been proposed recently in
Ref.~\onlinecite{meir2002}.
Although the Hamiltonian, Eq.~(\ref{Rejec_k}), is
similar to the usual Anderson Hamiltonian~\cite{anderson61}, we stress
the important difference that the $kd$-hybridization term above
arises solely from the Coulomb interaction, whereas in the usual
Anderson case it comes primarily from one-electron interactions.
These have been completely eliminated above by solving the
one-electron problem exactly. The resulting hybridization term
contains the factor $n_{d\bar{\sigma}}$, and hence disappears when the
localized orbital is unoccupied. This reflects the fact that an
effective double-barrier structure and resonant bound state
occurs via Coulomb repulsion only because of the presence of a
localized electron.

\begin{figure}[htb]
{\par\centering \resizebox*{0.9\columnwidth}{!}{\rotatebox{0}
{\includegraphics{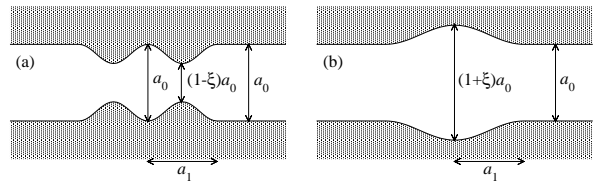}}} \par}
\caption{\label{oblika} The geometry of the 'open quantum dot' for
the parameterization (a) Eq.~(\ref{oblika2}) and (b) Eq.~(\ref{oblika1}).}
\end{figure}

To be specific, we consider in this paper a cylindrically symmetric
quantum wire with symmetry axis $z$ and lateral coordinates $r$
and $\varphi$
\cite{rejec003d}. Such a geometry corresponds to narrow 'v'-groove
$z$-dependent quantum
wires investigated recently, e.g. in Ref.~\onlinecite{kaufman99}. The diameter of the
wire is  \( a\left( z\right) \), with zero potential within the wire and
constant \(V_{0}>0 \) outside, i.e.,
\begin{equation}
\label{pot3d}
V\left( r,z\right) =\left\{ \begin{array}{ll}
0, & r<\frac{1}{2}a\left( z\right) \\
V_{0}, & r>\frac{1}{2}a\left( z\right) 
\end{array}.\right.
\end{equation}
For the wire width, two generic shapes are taken, shown in
Fig.~\ref{oblika} with

\begin{subequations}
\begin{eqnarray}
a\left( z\right)  & = & \left\{ \begin{array}{ll}
a_{0}\left( 1-\xi \sin ^{2}\pi \frac{z}{a_{1}}\right) , & \left| z\right| <a_{1}\\
a_{0}, & \left| z\right| >a_{1}
\end{array}\right. ,\label{oblika2} \\
a\left( z\right)  & = & \left\{ \begin{array}{ll}
a_{0}\left( 1+\xi \cos ^{2}\frac{\pi }{2}\frac{z}{a_{1}}\right) , & 
\left| z\right| <a_{1}\\
a_{0}, & \left| z\right| >a_{1}
\end{array}\right. .\label{oblika1} 
\end{eqnarray}
\end{subequations}

The region of interest is around \( z=0 \) and for large \( \left|
z\right| > a_1 \) the diameter is constant $a_0$.  Single particle
solutions corresponding to this geometry as well as the derivation and
calculation of parameters of the corresponding Hubbard Hamiltonian are
presented in Appendix A.

\section{Two-electron solutions}

\subsection{Bound states}

In order to calculate conductance though the system we first solve
the two interacting electron problem for the present geometry using the
extended Hubbard Hamiltonian Eq.~(\ref{hamdis1d}).
Solutions for bound states are determined
by numerical diagonalization of the system of 
equations presented in Appendix B,
Eq.~(\ref{sistem2el}). In Fig.~\ref{2dens} is shown the result 
of the two body electron density as 
a function of $z/a_0$ for various shapes of the bulge [Fig. \ref{oblika}(b)]. 
A general tendency is that long/narrow bulges correspond to stronger
interaction resulting in formation of a double peak in density, as
known from other studies of one-dimensional quantum dots \cite{jauregui}. As
long as the two peaks are not well separated, the
approximate methods mentioned
below are excellent, becoming
gradually less reliable with increasing separation between 
the peaks.

\begin{figure}[htb]
{\par\centering \resizebox*{0.9\columnwidth}{!}{\rotatebox{0}
{\includegraphics{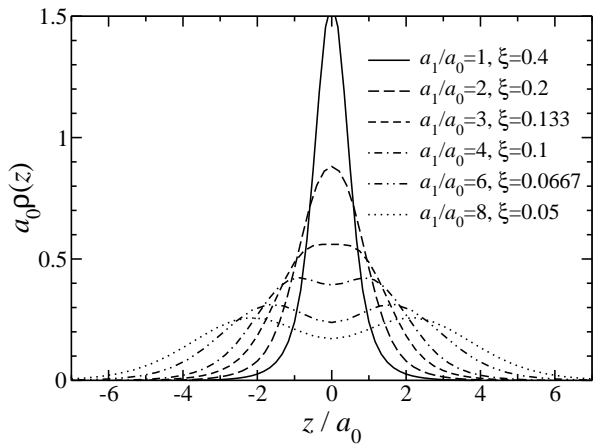}}} \par}
\caption{\label{2dens}Two-electron density for various bulge parameters.}
\end{figure}

\begin{figure}[htb]
{\par\centering \resizebox*{0.9\columnwidth}{!}{\rotatebox{0}
{\includegraphics{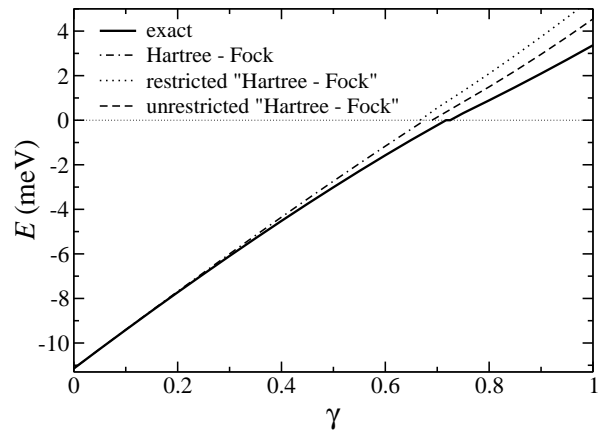}}} \par}
\caption{\label{1_24} Position of bound states ($E<0$) and resonances
($E>0$) vs. coupling $\gamma$, calculated exactly and within the
Hartree-Fock approximation. Wire
shape corresponds to Eq.~(\ref{oblika1}) with parameters: 
\protect\( a_{0}=a_{1}=10\textrm{ nm}\protect \), \protect\( \xi =0.24\protect \),
\protect\( V_{0}=0.4\textrm{ eV}\protect \) and \protect\( \kappa =50\textrm{ nm}
\protect \).}
\end{figure}

\begin{figure}[htb]
{\par\centering \resizebox*{0.9\columnwidth}{!}{\rotatebox{0}
{\includegraphics{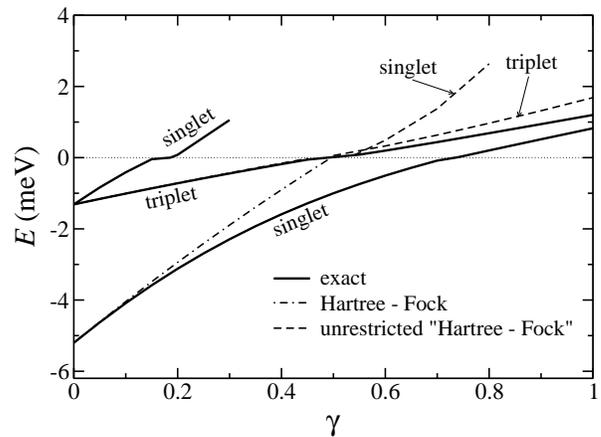}}} \par}
\caption{\label{1_06} As in Fig.~\ref{1_24} but with parameters:
\protect\( a_{0}=10\textrm{ nm}\protect \), \protect\( a_{1}=4a_{0}\protect \),
\protect\( \xi =0.06\protect \), \protect\( V_{0}=0.4\textrm{ eV}\protect \)
and \protect\( \kappa =50\textrm{ nm}\protect \).}
\end{figure}

In Fig.~\ref{1_24} we present typical examples of the energy of two
bound (singlet) electrons ($E<0$) where $\gamma$ is the electron-electron
coupling
strength, defined by replacement  $U\to
\gamma U$. Exact results are represented by the solid line, 
with other lines
representing results obtained with the
Hartree-Fock approximation, derived in Appendix C. At $E>0$ the lines
correspond to the position of the singlet resonance, calculated with
different methods and discussed below. In Fig.~\ref{1_06} the bulge is
longer and narrower, therefore both singlet and triplet bound states
exist for small $\gamma$, while for stronger coupling the triplet is
first pushed into continuum and finally, for $\gamma \sim 0.7$, both
states become resonances. Here approximate solutions are less
accurate, because the bulge is much larger than in the previous case
and therefore the problem is closer to the strong interaction limit,
as is seen also in Fig.~\ref{2dens}, dashed-dotted line, where the
dip in the electron density signals the strong interaction
regime. The Hartree-Fock approximation gives too large energies here,
which are, however, qualitatively correct.

\subsection{Scattering states}

Here we consider the scattering of an asymptotically free electron on
a bound electron within the bulge.  Such a system may be regarded as
an ``open quantum dot'' in which one electron is bound and inhibits
the transport of conduction electrons via Coulomb repulsion.  The
problem is analogous to treating the collision of an electron with a
hydrogen atom as, e.g., described in Ref.~\onlinecite{landauQM} and
studied by J.R. Oppenheimer and N.F. Mott\cite{oppenheimer28}.  We
only consider here cases in which the energy of the scattered electron
is smaller than the binding energy of the bound electron. This ensures
that only elastic scattering is possible. Asymptotically the two-body
wave function is a properly symmetrized product of a single particle
bound state, \(\left|\varphi\right>\), and scattered state, \( \left|
\chi \left( E\right) \right\rangle \).

For two electrons, the antisymmetrized wavefunction can be written as a
product of a spin-part and an orbital part. We write the orbital part as
\begin{equation}
\widetilde{\psi}_{ij}=\frac{\psi_{ij}+\left(-1\right)^{S}\psi_{ji}}{\sqrt{2}},
\end{equation}
ensuring that this is symmetric for singlets $(S=0)$ and antisymmetric for
triplets $(S=1)$ (see Appendix B).
For some large $N\gg1$, $\psi_{ij}$ takes the form
\begin{equation}
\label{nastavek}
\psi_{ij}=
\left\{\begin{array}{ll}\chi_{i}\left(E\right)\varphi_{j}
+r^{\left(S\right)}\chi_{i}^{\ast}\left(E\right)\varphi_{j},
& i<-N\\ t^{\left( S\right)}\chi _{i}\left(E\right)\varphi_{j}, & i>N
\end{array}\right. .
\end{equation}

Asymptotic solutions for the unbound electron are obtained from the
single-electron Hamiltonian Eq.~(\ref{ham1dis1d}) with the potential
\begin{equation}
\label{asimppot}
\tilde{\epsilon }_{j}=\epsilon _{j}+\sum _{k}U_{jk}\left| \varphi _{k}\right| ^{2}
\end{equation}
for large $|j|$. Here \( \left| \varphi \right\rangle \) is the
single-particle bound state in the potential \( \epsilon _{j} \).
Solutions with forward and backward currents have the following
asymptotic form $(j \to \infty)$
\begin{equation}
\label{asimpinf}
\chi _{j}=\left\{ \begin{array}{ll}
e^{\pm ikj} & \kappa <\infty \\
e^{\pm i\left( kj-\eta(k) \ln 2kj\right) } & \kappa =\infty 
\end{array}\right. ,
\end{equation}
for finite and infinite screening length $\kappa$, respectively (see
Appendix A). With no screening $(\kappa =\infty)$ $\chi _{j}$ are the
Coulomb functions \cite{abramovitz}.

Numerically exact solutions are obtained by solving a set of linear
equations for the \( \left( 2N+1\right) ^{2} \) variables \(
\widetilde{\psi }_{ij} \) and transmission and reflection amplitudes.

\section{Conductance}

\subsection{Single-electron solutions}

From the solution of the scattering problem, the conductance at zero
temperature is calculated using the usual Landauer-B\"{u}ttiker
formalism \cite{landauer57},
\begin{equation}
G=G_0 {\cal T}(E), \label{landauer}
\end{equation}
where $G_{0}=2e^{2}/h$, $E$ is the Fermi energy (in this case $E=E_1$) 
and ${\cal T}(E)$ is the 
total transmission probability. 

\begin{figure}[htb]
{\par\centering\resizebox*{0.9\columnwidth}{!}{\includegraphics{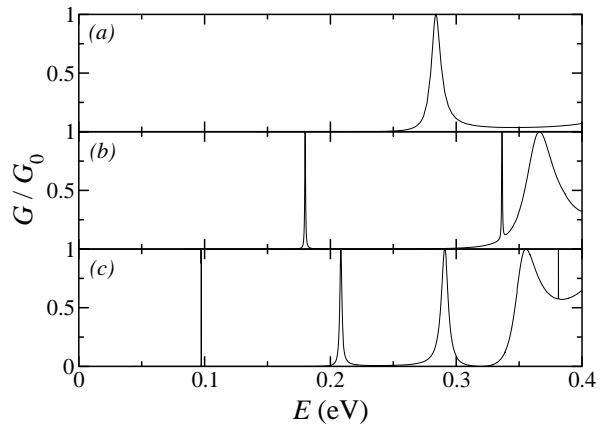}} \par}
\caption{\label{exact} $G$ for a wire with shape  Eq.~(\ref{oblika2}) and
\protect\( V_{0}=0.4\textrm{ eV}\protect \), \protect\( \xi =0.8\protect \),
\protect\( a_{1}=a_{0}\protect \) and in particular (a)  \protect\(
a_{0}=7\textrm{ nm}\protect \), (b) \protect\( a_{0}=10\textrm{
nm}\protect \), and (c) \protect\( a_{0}=15\protect \)
nm.}
\end{figure}

\begin{figure}[htb]
{\par\centering \resizebox*{0.9\columnwidth}{!}{\includegraphics{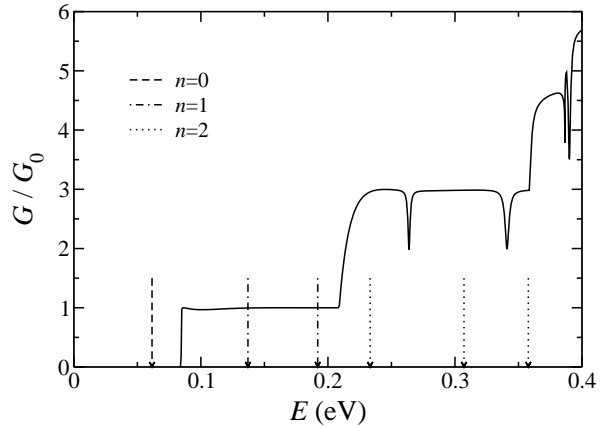}} \par}
\caption{\label{exactb}
$G$ for wire shape    Eq.~(\ref{oblika1}) and
\protect\( V_{0}=0.4\protect \) eV, \protect\( \xi =0.5\protect \),
\protect\( a_{0}=a_{1}=10\textrm{ nm}\protect \). Vertical lines
indicate positions of bound states for the lowest channels.}
\end{figure}

For an open bulge of shape Eq.~(\ref{oblika2}), Fig.~\ref{oblika}(a),
there are no bound states and only single-electron
solutions\cite{ramsak98} are relevant.  
In Fig.~\ref{exact} we present $G$ as a function of
electron energy for wires with shape Eq.~(\ref{oblika2}) and three different
widths. The main effect is a change of energy scale, according to
scaling rule Eq.~(\ref{transform}), and the magnitude of the conductance at 
the resonance energy, $G_0$. 
In Fig.~\ref{exactb} the conductance through the bulge of the 
shape from Eq.~(\ref{oblika1}) is presented. In contrast to the previous figure,
a bound state can exist here, indicated by the dashed vertical line.
Further lines ($n=1,2$) indicate bound
states of {\it individual channels} for the special case when channel
mixing terms in Eq.~(\ref{schrodkvazi1d}) are set to zero. Dips in the
conductivity in the second plateau ($G\sim 3 G_0$) correspond to Fano
resonances caused by interchannel mixing terms. 

\begin{figure}[htb]
{\par\centering \resizebox*{0.9\columnwidth}{!}{\rotatebox{0}
{\includegraphics{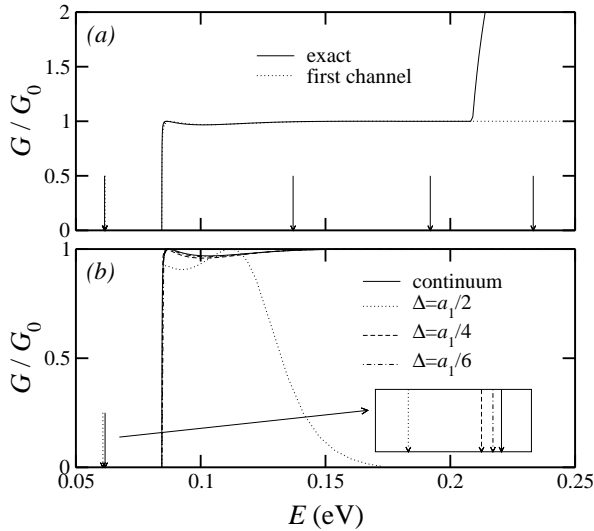}}} \par}
\caption{\label{2ch}(a) Conductance from Fig.~\ref{exactb} in
comparison with the result obtained in the one channel
approximation. (b) Conductance calculated with different \protect\(
\Delta \protect \) for wire parameters as in (a).  Bound states are
presented in the inset with enlarged energy scale.}
\end{figure}

In Fig.~\ref{2ch}(a) we again show the result of Fig.~\ref{exactb}
comparing it with the one-channel approximation. In this paper we
are interested in the rising edge of the conductance at the
threshold, but with Coulomb interactions between bound and
scattered electron included. Near threshold the
one-channel approximation is
excellent and therefore in the following we neglect higher channels.
In Fig.~\ref{2ch}(b) is presented the influence of discretization
parameter \protect\( \Delta \protect \), as introduced in Appendix A,
on the conductivity. The position of the bound state is not strongly
dependent on \protect\( \Delta \protect \) (inset in enlarged energy
scale) and for $\Delta < a_1/5$ the results obtained on the lattice
agree with the continuum calculation within a percent, which
justifies the use of the discretized Hamiltonian, Eq.~(\ref{hamdis1d}).
  
\subsection{\label{nepel}Interacting electrons}

We may extend the formula Eq.~(\ref{landauer}) to the case described in the
previous section in which one electron is bound in the wire and the remaining
electrons are transmitted with energy-dependent probability.
Let $P_{\sigma }$ be the probability that the bound electron has
spin $\sigma $. It follows directly that the conductance due to all spin-up
electrons in the leads is given by the extended Landauer B\"{u}ttiker
formula: 
\begin{equation}
G_{\uparrow }=\frac{e^{2}}{h} \left[ P_{\uparrow }{\cal T}_{\uparrow
\uparrow }(E )+P_{\downarrow }{\cal T}_{\uparrow \downarrow
}(E )\right],   \label{aa}
\end{equation}
where ${\cal T}_{\uparrow \uparrow }$ is the transmission probability when
the bound electron is spin up,  ${\cal T}_{\uparrow \downarrow }$\ is the
transmission probability when the bound electron is spin down and $E$ is the
Fermi energy. We have a
similar expression for spin-down electrons in the leads and hence the total
conductance is 
\begin{equation}
G(E)=\frac{e^{2}}{h} \left[ P_{\uparrow }{\cal T}_{\uparrow \uparrow
}(E )+P_{\downarrow }{\cal T}_{\uparrow \downarrow }(E
)+P_{\uparrow }{\cal T}_{\downarrow \uparrow }(E )+P_{\downarrow }%
{\cal T}_{\downarrow \downarrow }(E )\right] .
\label{bb}
\end{equation}
The transition probabilities ${\cal T}_{\uparrow \uparrow }$ and ${\cal T}%
_{\uparrow \downarrow }$ are different since in the former case the
conduction and bound electrons both have the same spin (up) before and after
scattering whereas in the latter case there are two possible final states,
with or without spin-flip, i.e. 
\begin{equation}
{\cal T}_{\uparrow \downarrow }(E)=
\left| t_{\uparrow\downarrow\rightarrow\uparrow\downarrow}\right|
^{2}+\left| t_{\uparrow\downarrow\rightarrow\downarrow\uparrow}\right|
^{2}, \label{tud}
\end{equation}
where the scattering amplitudes are defined by
\begin{eqnarray}
\langle i \uparrow j \downarrow | \psi_{\uparrow \downarrow}\rangle &\rightarrow&
t_{\uparrow\downarrow\rightarrow\uparrow\downarrow} \chi_i \varphi_j \\ \nonumber
\langle i \downarrow j \uparrow | \psi_{\uparrow \downarrow}\rangle &\rightarrow&
t_{\uparrow\downarrow\rightarrow\downarrow\uparrow} \chi_i \varphi_j 
\label{ijud}
\end{eqnarray}
as $i\to \infty$. $| \psi_{\uparrow \downarrow}\rangle$ is the exact 
scattering wavefunction and $|i \sigma j \sigma'\rangle=c^\dagger_{i 
\sigma} c^\dagger_{j \sigma'}  
|0 \rangle $. $\varphi_j=\langle j | \varphi \rangle$ is the bound-state one-electron 
wavefunction and $\chi_j=\langle j | \chi \rangle$ is a forward propagating one-electron
wavefunction at large $j$, as discussed in Section III.

In zero magnetic field it is clear that $P_\uparrow=P_\downarrow=\frac{1}{2}$ in
Eq.~(\ref{bb}). We can express $G(E)$ in a simpler form since the 
${\cal T}_{\sigma \sigma'}(E)$ are not all independent. Transforming to singlet 
and triplet base states (with $S_z=0$),
\begin{eqnarray}
|s,i,j\rangle &=&\frac{|i\uparrow ,j\downarrow \rangle -|i\downarrow ,j\uparrow
\rangle }{\sqrt{2}},\\ \nonumber
|t,i,j\rangle &=&\frac{|i\uparrow ,j\downarrow \rangle
+|i\downarrow ,j\uparrow \rangle }{\sqrt{2}},
\label{stij}
\end{eqnarray}
we get
\begin{eqnarray}
\langle s,i,j| \psi_{\uparrow \downarrow}\rangle &\rightarrow&
\frac{t^{(0)}}{\sqrt{2}} \chi_i \varphi_j \\ \nonumber
\langle t,i,j | \psi_{\uparrow \downarrow}\rangle &\rightarrow&
\frac{t^{(1)}}{\sqrt{2}} \chi_i \varphi_j, 
\label{stud}
\end{eqnarray}
where
\begin{equation}
\label{tupdown}
\begin{array}{rcl}
t^{(0) } &=&
t_{\uparrow\downarrow\rightarrow\uparrow\downarrow
}+t_{\uparrow\downarrow\rightarrow\downarrow\uparrow }  \\ 
 t^{(1) }&=&
t_{\uparrow\downarrow\rightarrow\uparrow\downarrow
}-t_{\uparrow\downarrow\rightarrow\downarrow\uparrow }
\end{array}.
\end{equation}
Hence, Eq.~(\ref{aa}) becomes\cite{landauQM} 
\begin{equation}
G=G_0\left(\frac{1}{4}\left| t^{\left( 0\right) }\right|
^{2}+\frac{3}{4} \left| t^{\left( 1\right) }\right| ^{2}\right).
\end{equation}

\subsection{Results of numerical analysis}

\begin{figure}[htb] {\par\centering
\resizebox*{0.9\columnwidth}{!}{\rotatebox{0}{\includegraphics{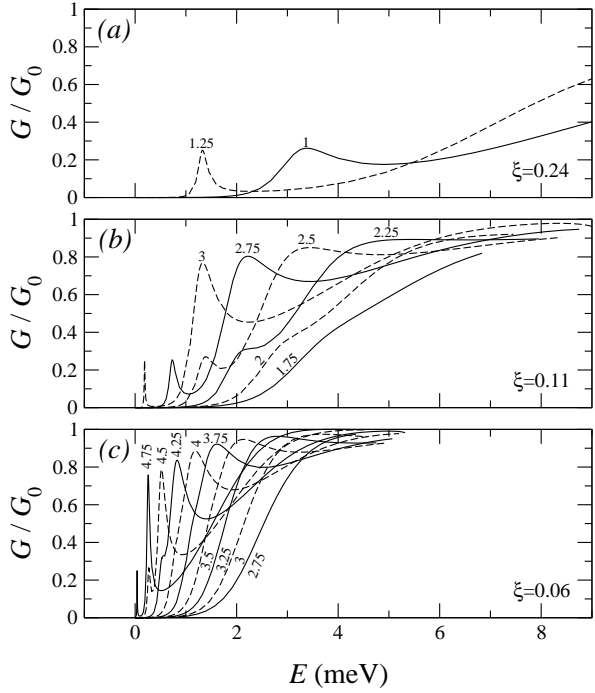}}}
\par}
\caption{\label{final1} Conductance for different shapes of the
bulge. Each line is labeled with the parameter
\protect\(a_{1}/a_{0}\protect \). Other wire parameters:
\protect\( a_{0}=10\textrm{ nm}\protect \),
\protect\( V_{0}=0.4\textrm{ eV}\protect \) and \protect\( \kappa
=50\textrm{ nm}\protect \).}
\end{figure}

In Fig.~\ref{final1} we present the result of a comprehensive study
of conductance for a variety of shapes of bulge. In
Fig.~\ref{final1}(a) the bulge is wide and so short that only a singlet
resonance is developed. The conductance therefore exhibits a structure
similar to the 0.3 anomaly found in experiment \cite{patel91}. In
Figs.~\ref{final1}(b,c) both singlet and triplet resonances are visible
with a tendency for the resonances to sharpen as the bulge becomes weaker
($\xi \to 0$). 

\begin{figure}[htb]
{\par\centering \resizebox*{0.9\columnwidth}{!}
{\rotatebox{0}{\includegraphics{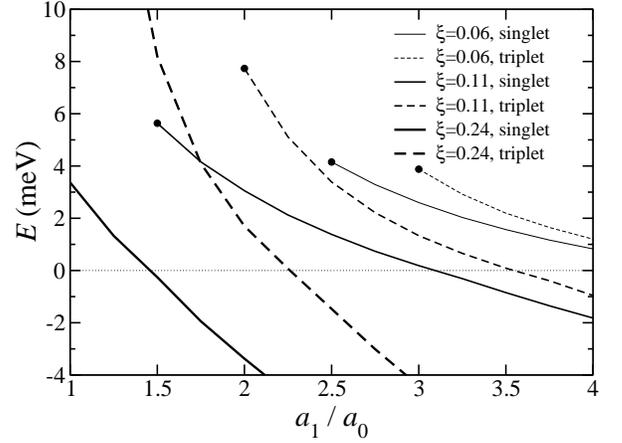}}} \par}
\caption{\label{final2} Energies of singlet (full lines) and triplet
(dashed lines) resonances and bound states for wire Eq.~(\ref{oblika1}) as
a function of \protect\( a_{1}/a_0\protect \) for different  \protect\( \xi \protect \).
Other parameters are as in Fig.~\ref{final1}. 
Full circles represent the energy, where
the resonance energy is above the 'ionization' energy and the underlying
wave function form is not valid anymore.}
\end{figure}

\begin{figure}[htb]
{\par\centering \resizebox*{0.9\columnwidth}{!}{\rotatebox{0}
{\includegraphics{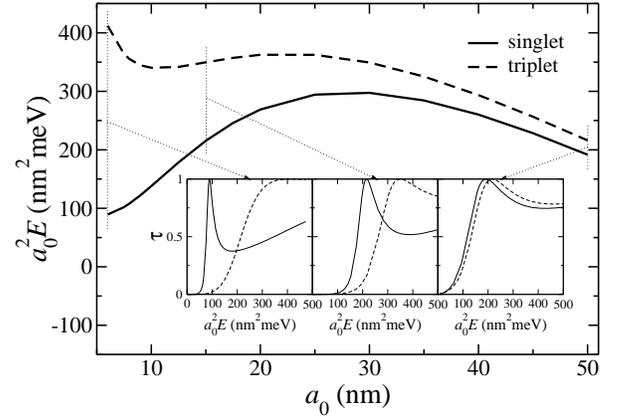}}} \par}
\caption{\label{final3}The position of singlet (full line) and triplet
(dashed line) resonances as a function of the width of the wire,
$a_0$. Note that the energy of the resonances is presented in a scaled
form. $a_1=50\mathrm{~nm}$, $\xi=0.11$ and other parameters are as in
Fig.~\ref{final1}.}
\end{figure}

In Fig.~\ref{final2} the wire width is fixed at $10$nm of the wire and
positions of singlet (full lines) and triplet (dashed lines)
resonances (or the corresponding bound states for $E<0$) are plotted
for various lengths and widths of bulge, represented by $a_1/a_0$ and
$\xi$. We see that the resonances survive for a wide range of
parameters. In Fig.~\ref{final3} is shown the position of singlet and
triplet resonance energies vs. the width of the wire, $a_0$, with
fixed shape of the bulge. The insets show the energy dependence of
singlet and triplet transmission probabilities for selected special
cases.  Note that we have scaled the energy by a factor $a_0^2
E$. This would produce identical curves for non-interacting electrons
[Eq.~(\ref{transform})].

After performing calculations for a wide range of parameters, we 
conclude that a singlet resonance is always lower in energy than 
its corresponding
triplet, in accordance with Lieb-Mattis theorem, which however, is
strictly valid only for ground states \cite{lieb62}.

\subsection{Magnetic field along the symmetry axis}

The nature of conductance anomalies studied here can be further
illuminated with experiments done in a strong magnetic
field\cite{patel91}. The effect of the magnetic field
is, in our treatment, taken into account via the usual Zeeman
splitting of channel energies. The incoming electron with Fermi
energy \( E \) and spin component \( S_{z}=\pm\frac{1}{2} \) then has
kinetic energy
\begin{equation}
E_{\textrm{k}}\equiv E_\mp =E\mp E_{\textrm{B}},
\end{equation}
where $E_{\rm B}=\frac{1}{2}g^\ast\mu_{\rm B} B$. $g^\ast$ is the effective gyromagnetic ratio and \( \mu
_{\textrm{B}} \) is Bohr magneton.

Near the conductance threshold we assume that the current is sufficiently low
that the localized electron is in its ground-state with spin $\downarrow$ before 
each scattering event with a conduction electron. Hence $P_\downarrow=1$ and
$P_\uparrow=0$ in Eq.~(\ref{aa}) which becomes
\begin{equation}
G_\uparrow(E,B)=\frac{e^{2}}{h} {\cal T}_{\uparrow \downarrow}(E,B).
\end{equation}
In this case $G_\downarrow \neq G_\uparrow$ but, rather,
\begin{equation}
G_\downarrow(E,B)=\frac{e^{2}}{h} {\cal T}_{\downarrow \downarrow}(E,B).
\end{equation}
Since ${\cal T}_{\uparrow \downarrow}(E,B)={\cal T}_{\uparrow \downarrow}(E_-,0)$ and
${\cal T}_{\downarrow \downarrow}(E,B)={\cal T}_{\downarrow \downarrow}(E_+,0)$,
then the conductance is
\begin{eqnarray}
G&=&G_{\uparrow}+G_{\downarrow}\\ \nonumber
&=&\frac{e^{2}}{h}[
{\cal T}_{\uparrow \downarrow}(E_-,0)+
{\cal T}_{\downarrow \downarrow}(E_+,0)] \\ \nonumber
&=& \frac{e^2}{h}[
{\cal T}_{t}(E_+,0)+
\frac{1}{2}{\cal T}_{t}(E_-,0)+
\frac{1}{2}{\cal T}_{s}(E_-,0)],\label{gb}
\end{eqnarray}
where ${\cal T}_{s}$ and  ${\cal T}_{t}$ are the same functions 
as in zero magnetic field.

\begin{figure}[htb]
{\par\centering \resizebox*{0.9\columnwidth}{!}{\rotatebox{-0}
{\includegraphics{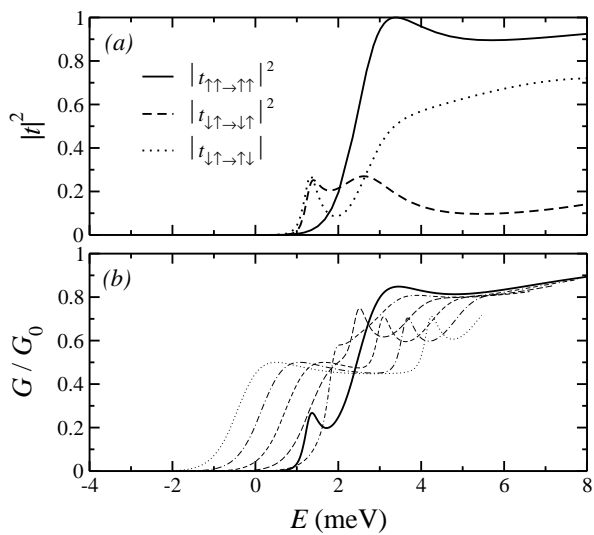}}} \par}
\caption{\label{magnet} (a) Transmission probabilities for relevant
spin configurations. (b) Conductance for $B=0$ (full line) and other
lines for $B$ in increments $\Delta B=10$~T. Parameters of the wire:
\protect\( a_{0}=10 \textrm{ nm}\protect \), \protect\(
a_{1}=2.5a_{0}\protect \), \protect\( \xi =1.11\protect \), \protect\(
V_{0}=0.4 \textrm{ eV}\protect \) and \protect\( \kappa
=50\textrm{ nm}\protect \).}
\end{figure}

In Fig.~\ref{magnet}(a) are plotted  individual transmission probabilities for
different spin configurations. Note that the spin-flip term
is in general dominant at higher energies. In Fig.~\ref{magnet}(b) the
corresponding results for conductance in the presence of a magnetic 
field is shown. The full line corresponds to $B=0$, and other curves to $B$ in 
increments $\Delta B=10$T. 

\subsection{Results for the Anderson model}

As shown in Section II, the Hubbard model studied above can be
mapped onto an extended Anderson model, Eq.~(\ref{Rejec_k}). 
Conductance through a quantum dot described by a standard Anderson model
is basically described by a peak or several peaks and at higher energies the
conductance approaches zero \cite{anddot}. In the case of an 
open quantum dot, studied here, 
at higher energies the conductance tends toward unity, as a consequence of
additional coupling parameters in the extended 
model. Here we analyze these terms individually
and show their relative importance.

\begin{figure}[tbh]
{\par\centering \resizebox*{0.7\columnwidth}{!}{\rotatebox{0}
{\includegraphics{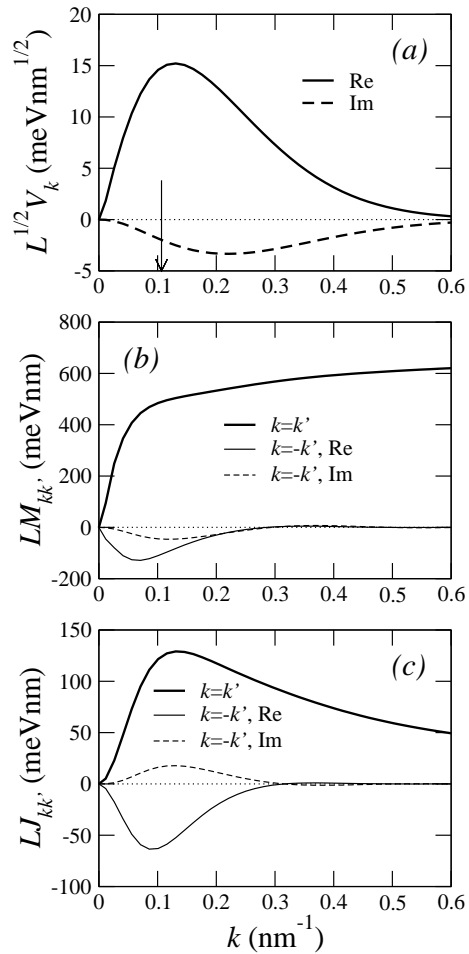}}} \par}
\caption{$k$-dependence of matrix elements of the extended Anderson
model. The wire is parameterized with $ a_{0}=10$~nm, $\protect\xi
=0.24$, $a_{1}/a_{0}=2$, $V_{0}=0.4$~eV, $\protect \kappa=50$~nm and
$\protect\gamma =1$. (a) Mixing coupling $V_{k}$. The energy
$\epsilon_d + U$ is indicated with an arrow. (b, c) Scattering
couplings $M_{kk'}$ and $J_{kk'}$. $L$ is the length of the wire, where the wave
functions are normalized.}
\label{Rejec_Fig1}
\end{figure}

The coupling parameters are momentum dependent and in 
Fig.~\ref{Rejec_Fig1} the couplings $V_{k}$, $M_{kk^\prime}$ and
$J_{kk^\prime}$ are shown. Note that $M_{kk^\prime}$ at higher
energies tends to a constant, while other parameters approach zero,
ensuring the correct behavior at high-energy with unit
transmission.
The scattering solutions of the Hamiltonian Eq. (\ref{Rejec_k}) are then
obtained exactly for two electrons with the boundary condition
that for $z\rightarrow \infty $, one electron occupies the lowest
bound state, whilst the other is in a forward propagating plane
wave state, $\phi _{k}(z)\sim e^{ikz}.$ From these solutions we
compute the conductance again using the Landauer-B\"{u}ttiker formula.

\begin{figure}[tbh]
{\par\centering \resizebox*{0.8\columnwidth}{!}{\rotatebox{0}
{\includegraphics{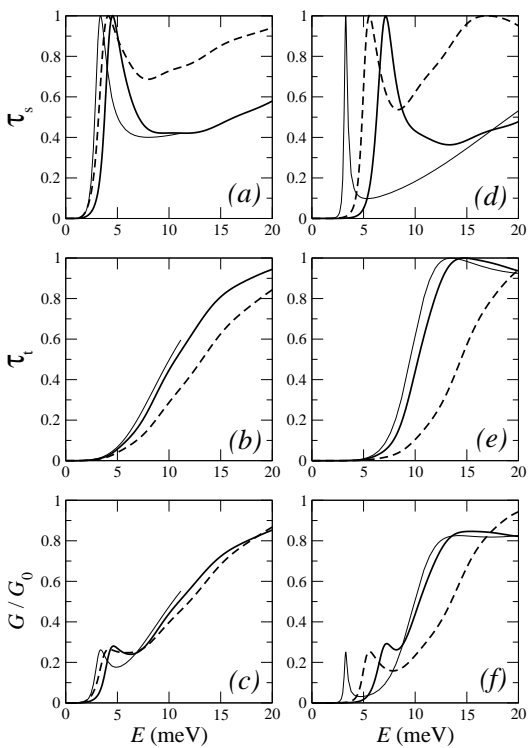}}} \par}
\caption{Singlet (a, d) and triplet (b, e) transmission probabilities and
corresponding conductances (c, f). Parameters for the left set are as in
Fig.~\ref{Rejec_Fig1}, for the right set: $a_{0}=10$~nm, $\protect\xi =0.15$,
$a_{1}/a_{0}=4$, $V_{0}=0.4$~eV, $\protect\kappa=100$~nm and
$\protect\gamma =0.9$.  Thin lines represent exact results from
Eq.~(\ref{hamdis1d}), thick lines are results from
Eq.~(\ref{Rejec_k}). Dashed lines show results where the exchange
term in Eq.~(\ref{Rejec_k}) is neglected.} \label{Rejec_Fig2}
\end{figure}

In Figs.~\ref{Rejec_Fig2}(a,b,c) we compare the results of
${\cal T}_{ \mathrm{s}}$,
${\cal T}_{\mathrm{t}}$ and conductance $G$ for a wire with the bulge
as in Fig.~\ref{1_24}.
The thin lines are the exact scattering result for two electrons. The
solid lines show the exact scattering solutions for the Anderson-type
Hamiltonian, for which the matrix elements, and their energy
dependence are calculated explicitly. The solution of this
Anderson-type model for two electrons, in which the localized level
always contains at least one electron, reproduce the main features of
the exact scattering solutions of the original model. The energy
dependence of the matrix elements is essential to get this good
agreement. Figs.~\ref{Rejec_Fig2}(d,e,f) show the corresponding results for
a longer bulge from Fig.~\ref{1_06}. Also shown in  Fig.~\ref{Rejec_Fig2}, dashed
lines, are results with the direct exchange term omitted in 
Eq.~(\ref{Rejec_k}). This term can have a significant quantitative effect,
but does not qualitatively change the conductance curves.

We have also solved a similar model in which plane waves,
rather than exact scattering states of the non-interacting problem,
were used. However, this gave poor agreement with the exact results.
We conclude that an Anderson-type model is adequate for a near-perfect
quantum wire provided that a suitable basis set is used and the
energy-dependence of the matrix elements is accurately
determined. Future work will focus on the many-electron properties of
this effective Hamiltonian, including `Kondo' and `mixed valence'
regimes.

\subsection{Approximate methods}

It is not easy to get sufficiently accurate numerical solutions
for the case of more
than two electrons. Therefore it would be extremely useful if an
accurate
approximative method could be applied. The simplest approximation
(presented here for the case of two electrons) is be the first iteration in
solving the Hartree-Fock equations, as we presented in Section III for 
the case of bound
states.

We assume the two-body wave functions consist of a single particle
state \( \left| \varphi ^{\left( 1\right) }\right\rangle \) and
scattering state $ | \chi \rangle$ with energy \( E \). The
two-electron wavefunction has then the form (see Appendix C),
\begin{equation}
\label{psi_approx_2}
\left| \psi \right\rangle =\sum _{ij}\varphi _{i}\chi _{j}c_{ij}^{
\left( S,S_{z}\right) \dagger }\left| 0\right\rangle .
\end{equation}
The coefficients \( \varphi _{i} \) are known, therefore 
only coefficients  \( \chi _{i} \) must be
determined. For the singlet state the simplest approximation is
obtained if we perform the first iteration of the Hartree-Fock method
subject to additional condition that the electron has energy $E$
(``restricted Hartree-Fock'' approximation): 
\begin{equation}
\label{rest_sing_tr}
\left\langle 0\left| c_{i}H_{1}\right| \chi \right\rangle +\sum
_{j}U_{ij}
\left| \varphi _{j}\right| ^{2}\chi _{i}=E\chi _{i},
\end{equation}
where, using Eq.~(\ref{ham1dis1d}), 
$\left\langle 0\left| c_{i}H_{1}\right| \chi \right\rangle=
-t(\chi_{i-1}+\chi_{i+1})+\epsilon_i \chi_i$. This is just the tight binding
results for a single electron moving in an effective potential
$\epsilon _{i}+\sum _{j}U_{ij}\left| 
\varphi _{j}\right| ^{2}$.

For the triplet state, the result is:
\begin{equation}
\label{trip_tr}
\left\langle 0\left| c_{i}H_{1}\right| \chi \right\rangle +
\sum _{j}U_{ij}\left| \varphi _{j}\right| ^{2}\chi _{i}-\sum _{j}
U_{ij}\varphi _{j}^{\ast }\varphi _{i}\chi _{j}=E\chi _{j}.
\end{equation}
A better approximation for the singlet case starts from the unrestricted
Hartree-Fock approximation, where the energy is  
\begin{eqnarray}
&&\left\langle \psi \left|H\right| \psi \right\rangle  =
 \left\langle \varphi \left| H_{1}\right| \varphi 
\right\rangle \left\langle \chi |\chi \right\rangle 
+
\left\langle \chi \left| H_{1}\right| \chi \right\rangle 
\left\langle \varphi |\varphi \right\rangle+ \nonumber \\
&&\quad\quad+\left( -1\right) ^{S}
\left( \left\langle \varphi \left| H_{1}\right| \chi \right\rangle 
\left\langle \chi |\varphi \right\rangle +\left\langle \chi \left| 
H_{1}\right| \varphi \right\rangle \left\langle \varphi |
\chi \right\rangle \right)+ \nonumber \\
&&\quad\quad+\frac{1}{2}\sum _{ij}U_{ij}\left| \varphi _{i}\chi _{j}+
\chi _{i}\varphi _{j}\right| ^{2},
\end{eqnarray}
and the norm is given with
\begin{equation}
\left\langle \psi |\psi \right\rangle =\sum _{i}\left| \chi _{i}
\right| ^{2}+\left| \sum _{i}\varphi _{i}^{\left( 1\right) \ast }\chi _{i}\right| ^{2}.
\end{equation}
The 
coefficients $\chi_i$ are calculated from the Hartree-Fock
equations based on the variation principle (``unrestricted
Hartree-Fock'' approximation),
\begin{eqnarray}
\label{unrest_sing_tr}
&&\left\langle 0\left| c_{i}H_{1}\right| \chi \right\rangle +
\sum _{j}U_{ij}\left| \varphi _{j}\right| ^{2}\chi _{i}+
\sum _{j}U_{ij}\varphi _{j}^{\ast }\varphi _{i}\chi _{j}+ \nonumber \\
&&\quad\quad+
\left( E-E^{\left( 1\right) }_{1}\right) \sum _{j}\varphi _{j}^{\ast }
\varphi _{i}\chi _{j}=E\chi _{i}.
\end{eqnarray}
    
\begin{figure}[htb]
{\par\centering \resizebox*{0.6\columnwidth}{!}{\rotatebox{0}
{\includegraphics{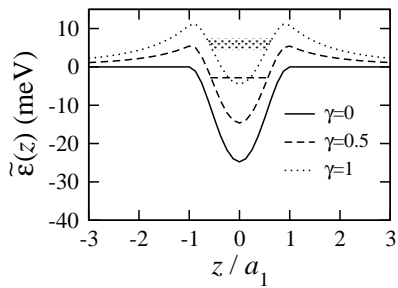}}} \par}
\caption{\label{efpot} Effective potential from
Eq.~(\ref{rest_sing_tr}) and for the wire with parameters as in Fig.~\ref{1_24}.}
\end{figure}

\begin{figure}[htb]
{\par\centering \resizebox*{0.9\columnwidth}{!}{\rotatebox{0}
{\includegraphics{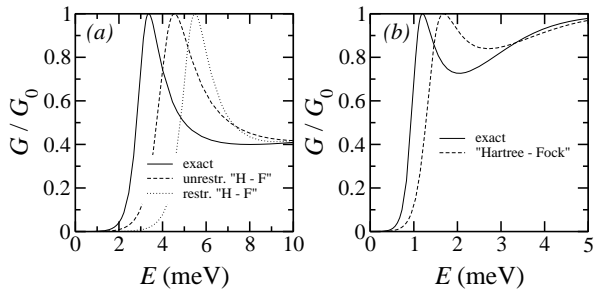}}} \par}
\caption{\label{approx}(a) Singlet resonance (parameters from
Fig.~\ref{1_24}). Exact result and approximations of
Eqns.~(\ref{rest_sing_tr}) and (\ref{unrest_sing_tr}) are shown. (b)
Triplet resonance (parameters from Fig.~\ref{1_06}). Exact result and
approximation of Eqn.~(\ref{trip_tr}) are shown.}
\end{figure}

In Fig.~\ref{efpot} the effective one
dimensional potential in Eq.~(\ref{rest_sing_tr}) is plotted for
$\gamma=0, 0.5$ and $1$. The shaded region
represents the position and the width of single particle resonance in
this effective potential. This resonance corresponds to the singlet
resonance presented in Fig.~\ref{approx}(a) (dashed line), for wire
parameters given in Fig.~\ref{1_24} together
with exact result, full line, and calculated from
Eq.~(\ref{rest_sing_tr}).  

We also show in Fig.~\ref{approx}(a) the exact result and the corresponding
result for an unrestricted Hartree-Fock scheme for which the wavefunctions
of up and down spin electrons are different. As expected, the unrestricted method
gives a more accurate result though both methods reproduce the main features,
the main discrepancy being an overall energy shift. Similarly in 
Fig.~\ref{approx}(b) we 
present the corresponding triplet resonance curve for parameters
from Fig.~\ref{1_06}. Again, the overall agreement with the exact result 
is good apart from an overall energy shift.

\section{Summary and conclusions}

We have shown that quantum wires with weak longitudinal
confinement, or open quantum dots, can give rise to spin-dependent, Coulomb
blockade resonances when a single electron is bound in the confined region.
This is a universal effect in one-dimensional systems with very weak
longitudinal confinement. The emergence of a specific structure at $G(E)\sim 
\frac{1}{4}\frac{2e^{2}}{h}$ and $G\sim \frac{3}{4}\frac{2e^{2}}{h}$ is a
consequence of the singlet and triplet nature of the resonances and the
probability ratio 1:3 for singlet and triplet scattering and as such is a
universal effect. A comprehensive numerical investigation of open quantum
dots using a wide range of parameters shows that singlet resonances are
always at lower energies than the triplets, in accordance with the
corresponding theorem for bound states \cite{lieb62}. With increasing
in-plane magnetic field, the resonances shift their position and eventually
merge in the conductance plateau at $G\sim e^{2}/h$. With increasing
source-drain bias we have shown why the higher triplet resonance weakens at
the expense of the singlet, with the latter surviving to the point where the
conductance steps themselves disappear.

\begin{figure}[htb]
{\par\centering \resizebox*{0.8\columnwidth}{!}{\rotatebox{0}
{\includegraphics{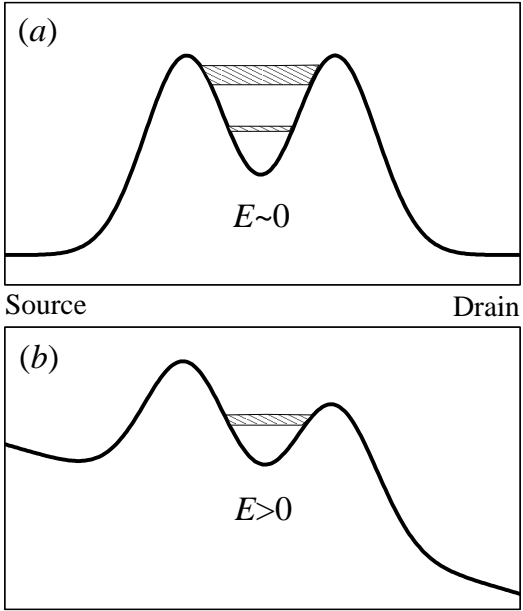}}} \par}
\caption{\label{land6}Effective double barrier showing singlet and triplet
resonance with very small source-drain bias (a) and large source-drain 
bias (b).}
\end{figure}

The existence of the conductance anomalies is a direct
consequence of an effective double-barrier potential seen by the conduction
electrons propagating from source to drain contacts under the influence of a
bound electron. For a symmetric one-electron confining potential, the
existence of a bound state is guaranteed but this is not necessarily the
case when the confinement is asymmetric. Such asymmetry in the confining
potential may be easily achieved under a finite source drain bias and
indeed, this was reported in some of the experiments on gated quantum wires 
\cite{thomas96,patel91}. These experiments show that as the source-drain
bias is increased from zero, an anomaly appears at\ $G\sim
0.25(2e^{2}/h)$, coexisting with the $0.7(2e^{2}/h)$ anomaly.  
Eventually, at larger bias, the remaining
anomaly also disappears but only when the conductance steps themselves are
on the point of disappearing, showing that the singlet anomaly is extremely
robust. This behaviour is consistent with our model since under bias the
triplet resonant bound-state will eventually disappear because the confining
potential in the $x$-direction will only accommodate a single one-electron
bound state, giving rise to a singlet resonance only. This is shown
schematically in Fig.~\ref{land6} where we also indicate the surviving singlet
becoming broader with increasing bias resulting in a more pronounced step, as
observed.

Finally, we speculate on the exciting possibility that these anomalies
in conduction are themselves a signature for a new kind of conducting
state in ultra clean wires close to the conduction threshold. Indeed,
there is some experimental evidence for this in that the anomalies
discussed above merge into a conductance step at $e^{2}/h$ under quite
moderate magnetic fields and in the cleanest samples this behaviour is
sometimes even seen in zero magnetic field. This suggests that there
may be an underlying spin polarized state associated with the
propagating electrons in the quasi 1D region. Such a spin-polarized
state would appear to violate the Lieb-Mattis theorem
\cite{lieb62} and
would also need to be made consistent with our above explanation in
terms of singlet and triplet resonances. In this respect we emphasize
that the above theory must break down at very low electron density in
the wire such that the mean separation between electrons in the wire
is somewhat greater than the effective Bohr radius, the so-called
strong correlation regime. In practical situations it is very
difficult to avoid some kind of weak potential fluctuation which traps
one electron. Indeed this may ultimately be impossible since even in a
nominally perfect wire, the presence of a single electron will
polarise its environment leading to a potential well which will bind
the electron giving rise to a Coulomb blockade for the remaining
electrons, though the energy scale (temperature) for this may be very low making
it susceptible to masking the other effects.

The main question is whether or not this confinement is
sufficiently large for the electron density to exceed to the inverse
Bohr radius when the wire begins to conduct. If the density remains
low at this conductance threshold then we cannot ignore the mutual
interaction between all electrons in the wire region, or even treat
them self-consistently. In this situation, a more appropriate picture
would be one in which the Coulomb repulsion dominates and maintains
roughly equal separation between the electrons as in a Wigner
chain. On the other hand, if the mean electron density is of order,
or greater,than the inverse Bohr radius, then an open quantum dot
picture with effective resonance levels for the propagating electron is
more appropriate, as discussed in this paper.

At low temperatures strong many-body effects are indicated from the
activation-like behaviour of the conductance \cite{thomas96,
kondo2002} and the thermopower coeficient \cite{appleyard00}. As
discused in our recend thermopower analysis, Ref.~\onlinecite{rejec02}, the
anomaly at low-temperatures may well be a many-body Kondo-like effect
contained within our extended Anderson model,
Eq.~(\ref{Rejec_k}), and studied recently in Ref.~\onlinecite{meir2002}, but
not within the two-electron approximation we have used here and in
some our earlier papers. It may well be that the two-electron
approximation breaks down at low temperatures.  The model presented
here differs from the standard Anderson model in that the
hybridisation term contains the factor $n_{-\sigma }$, and hence
disappears when the localized orbital is unoccupied. This reflects the
fact that an effective double-barrier structure and resonant bound
state occurs via Coulomb repulsion only because of the presence of a
localized electron. The standard results for the single impurity
problem \cite{hewsonbook} 
thus cannot be applied directly to this effective model,
and are a subject of current reaserch \cite{rejecram}.  However, a
Kondo-like resonance is expected \cite{boese01,meir2002}, for which
many-body effects would dominate with a breakdown of our two electron
approximation.

\appendix

\section{Cylindrical wire}

\subsection{Single electron basis}

A single electron in the wire considered here is described with the wave function 
\( \Psi (r,\varphi ,z) \), which is a solution of
the Schr\"{o}dinger equation

\begin{equation}
\label{schrod3d}
-\frac{\hbar ^{2}}{2m^{\ast }}\nabla ^{2}\Psi \left( r,\varphi
,z\right) 
+V\left( r,z\right) \Psi \left( r,\varphi ,z\right) =E\Psi \left(
r,\varphi 
,z\right) ,
\end{equation}
where the effects of nonparabolicity are neglected and the effective
mass is taken constant, $m^{\ast } = 0.067m_{\rm elec}$
with dielectric constant 12.5, appropriate
for GaAs \cite{ramsak98}.

At fixed \( z \) the wave function \( \Psi (r,\varphi ,z) \) is
expanded in a two-dimensional basis \( \Phi _{mn}\left( r,\varphi
;z\right) \) for the corresponding potential \( V\left( r;z\right) \)
, Eq.~(\ref{pot3d}). The coefficients in such an expansion over {\it
channels} are \( \psi _{mn}\left( z\right) \)
\begin{equation}
\Psi (r,\varphi ,z)=\sum _{n=0}^{\infty }\sum ^{n}_{m=-n}\psi
_{mn}\left( z\right) 
\Phi _{mn}\left( r,\varphi ;z\right) .\label{a2}
\end{equation}
The transverse wavefunftions, $\Phi _{mn}\left( r,\varphi ;z\right)$, depend
only parametrically on $z$ and take the form:

\begin{subequations}
\begin{equation}
\Phi _{mn}\left( r,\varphi ;z\right) = \left\{ \begin{array}{ll}
A_{mn}\left( z\right) J_{m}\left( k_{mn}\left( z\right) r\right) e^{im\varphi } & 
r<\frac{a(z)}{2} \\
\begin{array}{l}
\left[B_{mn}\left( z\right) B_{m}^{(1)}\left( \kappa _{mn}\left(
z\right) r\right)+\right.\\
\left.+C_{mn}\left( z\right) B_{m}^{(2)}\left( \kappa
_{mn}\left( z\right) r\right)\right]e^{im\varphi }
\end{array}& r>\frac{a(z)}{2}
\end{array}\right.,
\end{equation}
\begin{equation}
k_{mn}\left( z\right)  = \sqrt{\frac{2m^{\ast }\epsilon _{mn}\left( z\right) }
{\hbar ^{2}}},
\end{equation}
\begin{equation}
\kappa _{mn}\left( z\right)  = \sqrt{\frac{2m^{\ast }\left|
\epsilon _{mn}\left( z\right) 
-V_{0}\right| }{\hbar ^{2}}}.
\end{equation}
\end{subequations} 
Here 
\( B_{m}^{(1)}=I_{m} \)
and \( B_{m}^{(2)}= K_{m} \) are appropriate Bessel eigenfunctions \cite{abramovitz}
for $\epsilon_{mn} \leq V_0$ with
\( B_{m}^{(1)}=J_{m} \)
and \( B_{m}^{(2)}= Y_{m} \), for $\epsilon_{mn} > V_0$ . The
coefficients \( A_{mn} \), \( B_{mn} \),
\( C_{mn} \) and energies \( \epsilon _{mn} \) are determined from the
boundary conditions and the normalization of wave functions.

Substituting Eq.~(\ref{a2}) into Eq.~(\ref{schrod3d}) and integrating over
$r$ and $\varphi$ leads to following coupled ordinary differential
equations for $\psi_{mn}$,
\begin{eqnarray}
\label{schrodkvazi1d}
\psi ^{\prime \prime }_{mn}&+&\left[ k^{2}-k_{mn}^{2}\left( z\right)
+a_{mnn}\left( z\right) \right] 
\psi _{mn}+ \\ \nonumber
&+&\sum _{n\neq n^{\prime }}b_{mnn^{\prime }}(z)\psi ^{\prime
}_{mn^{\prime }}+
\sum _{n\neq n^{\prime }}a_{mnn^{\prime }}\left( z\right) \psi _{mn^{\prime }}=0,
\end{eqnarray}
where the coupling coefficients are

\begin{subequations}
\begin{equation}
a_{mnn^{\prime }}\left( z\right) = 2\pi \int _{0}^{R}\Phi
_{mn}^{\ast }\left( r,\varphi ;z\right) 
\frac{\partial ^{2}}{\partial z^{2}}\Phi _{mn^{\prime }}\left(
r,\varphi ;z\right) 
r\textrm{d}r,\label{amnn}
\end{equation}
\begin{equation}
b_{mnn^{\prime }}\left( z\right)  = 4\pi \int _{0}^{R}\Phi
_{mn}^{\ast }\left( r,\varphi ;z\right) 
\frac{\partial }{\partial z}\Phi _{mn^{\prime }}\left( r,\varphi ;z\right) r
\textrm{d}r.\label{bmnn} 
\end{equation}
\end{subequations}
The coefficients coupling channels with different \( m \) are 
zero due to the orthogonality of \( e^{im\varphi } \) for different
\( m \).

Note that the Schr\"{o}dinger equation, Eq.~(\ref{schrod3d}), is invariant under the
transformation

\begin{eqnarray}
\bf {r} & \rightarrow  & \Lambda\bf {r},\label{invar} \\
E,V & \rightarrow  & \Lambda^{-2}E,\Lambda^{-2}V.\label{transform}
\end{eqnarray}

\subsection{Extended Hubbard Hamiltonian}

We consider here the case when
the variation  in wire width is small, resulting in small derivatives of the
coefficients in Eqs.~\ref{amnn},
\ref{bmnn}. We consider only electrons with energy below the 
second channel and hence Eq.~(\ref{schrodkvazi1d}) reduces to a single equation for
motion in  $z$ direction, with the potential

\begin{equation}
\label{pot1d}
\epsilon \left( z\right) =\epsilon _{00}\left( a\left(z\right)\right) +a^{\prime
2}\left( z\right) 
\tilde{\epsilon }_{00}\left( a\left(z\right)\right) .
\end{equation}

The first term is the energy of the first channel and the second is
related to \( a_{000} ( z ) \) from
Eq.~(\ref{schrodkvazi1d}). $a^{\prime}(z)$ is the derivative of the
wire diameter with respect to $z$. The second term in Eq.~(\ref{pot1d})
is always positive since
 
\begin{equation}
\tilde{\epsilon }_{00}\left( a\right) =\frac{\hbar ^{2}\pi }{m^{\ast
}}\int _{0}^{R}\left( \frac{\partial \Phi _{00}\left( r,\varphi
\right) }{\partial a}\right) ^{2}r\textrm{d}r .
\end{equation}
The potential Eq.~(\ref{pot1d}) is  constant for large \( \left| z\right|  \)
and  set to zero for convenience, i.e.

\begin{equation}
\epsilon \left( z\right) \rightarrow \epsilon \left( z\right) -\epsilon 
\left( \infty \right).
\end{equation}
 
\begin{figure}[htb]
{\par\centering \resizebox*{0.7\columnwidth}{!}{\rotatebox{0}
{\includegraphics{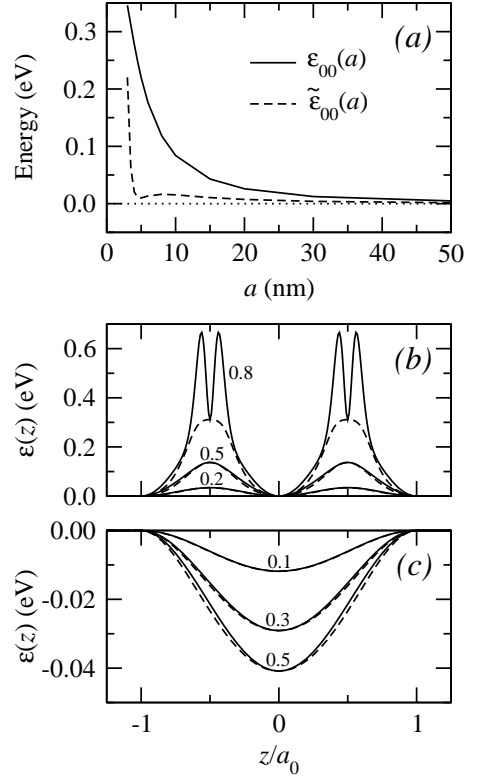}}} \par}

\caption{\label{epsil}(a) Dependence of \protect\( \epsilon _{00}\protect \) and
 \protect\( \tilde{\epsilon }_{00}\protect \)
in Eq.~(\ref{pot1d}) on wire diameter for  
\protect\( V_{0}=0.4\textrm{ eV}\protect \).
(b) One dimensional potential Eq.~(\ref{pot1d}) for the wire shape
Eq.~(\ref{oblika2}) and
various values of  \protect\( \xi \protect \). Dashed lines
correspond to the contribution  \protect\( \epsilon _{00}\protect
\). (c) The same as in (b) but for wire shape Eq.~(\ref{oblika1}).}
\end{figure}

\noindent In Fig.~\ref{epsil}(a) $\epsilon_{00}(a)$ and
$\tilde{\epsilon}_{00}(a)$ are presented as a function of wire
diameter. Figs.~\ref{epsil}(b) and (c) show the variation of
one-dimensional potential $\epsilon(z)$ along the wire.

\begin{figure}[htb]
{\par\centering \resizebox*{0.7\columnwidth}{!}{\rotatebox{0}
{\includegraphics{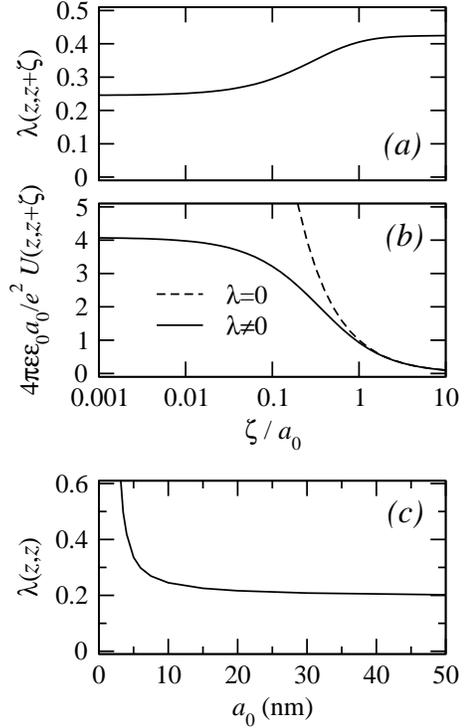}}} \par}

\caption{\label{lambde}
(a) Lateral distance vs. separation along the wire. (b)
Electron - electron interaction as a function of longitudinal
separation. In both cases is  \protect\( a_{0}=10\textrm{ nm}\protect
\) and \protect\( V_{0}=0.4\textrm{ eV}\protect \) and $a=$const. (c)
Lateral distance at fixed  \protect\( z\protect \) vs. wire diameter
at  \protect\( V_{0}=0.4\textrm{ eV}\protect \).}
\end{figure}

The single-electron Hamiltonian in the single channel approximation then becomes
\begin{equation}
H_1=-\frac{\hbar^2}{2m^\ast}\frac{\textrm{d}^{2}}{\textrm{d}z^{2}} +\epsilon(z).
\label{h1kont}
\end{equation}
This is readily generalized to many electrons,
\begin{eqnarray}
&&H = \sum _{\sigma }\int \psi _{\sigma }^{\dagger }\left( z\right)
\left[ 
-\frac{\hbar ^{2}}{2m^\ast}\frac{\textrm{d}^{2}}{\textrm{d}z^{2}} \right] \psi _{\sigma
}\left( z\right) 
\textrm{d}z+\\ \nonumber
&& +\frac{1}{2}\sum _{\sigma ,\sigma ^{\prime }}\int\!\!\! \int \psi
_{\sigma }^{\dagger }\left( z\right) 
\psi _{\sigma ^{\prime }}^{\dagger }\left( z^{\prime }\right) U\left(
z,z^{\prime }\right) 
\psi _{\sigma ^{\prime }}\left( z^{\prime }\right) \psi _{\sigma
}\left( z\right) 
\textrm{d}z\textrm{d}z^{\prime },\label{hamcont1d} 
\end{eqnarray}
where
\( \psi _{\sigma }^{\dagger }\left( z\right)  \) creates an electron with 
spin \( \sigma  \) at coordinate \( z \) and 
\begin{equation}
\label{uab}
U(z_i,z_j)=\frac{e^{2}}{4\pi \epsilon \epsilon _{0}d\left( z_{i},z_{j}\right) }
\end{equation}
with
\begin{equation}
\label{averdist}
\frac{1}{d\left( z_{i},z_{j}\right) }=\int \textrm{d}{\bf r}
_{i}\textrm{d}
{\bf r} _{j}\frac{\left| \Phi _{00}\left( {\bf r} _{i};z_{i}\right) 
\right| ^{2}\left| \Phi _{00}\left( {\bf r} _{j};z_{j}\right) 
\right| ^{2}}{\sqrt[]{\left( z_{i}-z_{j}\right) ^{2}+
\left| {\bf r} _{i}-{\bf r} _{j}\right| ^{2}}}.
\end{equation}
The Hamiltonian is further
discretized at points  \( z_{j}=j\Delta  \), new creation operators
are defined as

\begin{equation}
c_{j\sigma }^{\dagger }=\sqrt{\Delta }\psi _{\sigma }^{\dagger }\left( z_{j}\right) .
\end{equation}
For sufficiently small 
\( \Delta \) the difference formula is justified,

\begin{equation}
\left[ \frac{\textrm{d}^{2}}{\textrm{d}z^{2}}\psi _{\sigma }\left(
z\right) 
\right] _{z=z_{i}}\approx \frac{\psi _{\sigma }\left( z_{i-1}\right) -
2\psi _{\sigma }\left( z_{i}\right) +\psi _{\sigma }\left( z_{i+1}\right) }
{2\Delta ^{2}},
\end{equation}
and Eq.~(\ref{hamcont1d}) becomes the
discretized extended Hubbard Hamiltonian,
\begin{equation}
\label{hamdis1d1}
H=\sum _{\sigma }H_{1\sigma }+\frac{1}{2}\sum _{i\neq
j}U_{ij}n_{i}n_{j}+
\sum _{i}U_{ij}n_{i\uparrow }n_{j\downarrow },
\end{equation}
where  \( H_{1\sigma } \) is single-particle contribution for spin $\sigma$,
\begin{equation}
\label{ham1dis1d1}
H_{1\sigma }=-t\sum _{i}\left( c_{i+1\sigma }^{\dagger }c_{i\sigma
}+c_{i\sigma }^{\dagger }
c_{i+1\sigma }\right) +\sum _{i}\epsilon _{i}n_{i\sigma ,}
\end{equation}
with \( n_{i\sigma }=c_{i\sigma }^{\dagger }c_{i\sigma } \), 
\( n_{i}=\sum _{\sigma }n_{i\sigma } \), 
hoping parameter, 
\begin{equation}
\label{diskr_t}
t=\frac{\hbar ^{2}}{2m^{\ast }\Delta ^{2}},
\end{equation}
and $\epsilon _{i}=2t+\epsilon 
\left( z_{i}\right) $.
The effective distance between electrons at $z_i$ and $z_j$ is after
integrating Eq.~(\ref{uab}) over angular variables,
\begin{eqnarray}
&&\frac{1}{d\left( z_{i},z_{j}\right) }=8\pi \int _{0}^{R}r_{i}
\textrm{d}r_{i}\int _{0}^{R}r_{j}\textrm{d}r_{j}\left| \Phi _{00}
\left( r_{i};z_{i}\right) \right| ^{2}\times \nonumber\\
&&\quad\quad\times
\left| \Phi \left( r_{j};z_{j}
\right) \right| ^{2}\frac{\textrm{K}\left( -\frac{4r_{i}r_{j}}
{\left( z_{i}-z_{j}\right) ^{2}+\left( r_{i}-r_{j}\right) ^{2}}\right)}
{\sqrt[]{\left( z_{i}-z_{j}\right) ^{2}+\left( r_{i}-r_{j}\right) ^{2}}},
\end{eqnarray}
where $\rm K$ is the complete elliptic integral of the first kind. \( d\left(
z_{i},z_{j}\right)  \) can be decomposed into distance
along the wire and effective distance in the lateral direction, \(
\lambda \left( z_{i},z_{j}\right) a_{0} \), i.e. 

\begin{equation}
\frac{1}{d\left( z_{i},z_{j}\right) }=\frac{1}{\sqrt[]{\left(
z_{i}-z_{j}\right)^{2}+
\left[ \lambda \left( z_{i},z_{j}\right) a_{0}\right] ^{2}}}.
\end{equation}
The distance \( \lambda \left( z_{i},z_{j}\right)  \) is invariant under
the transformation Eq.~(\ref{invar}), and hence the potential,
Eq.~(\ref{uab}), transforms as
\begin{equation}
\label{invar_u}
U\rightarrow \Lambda^{-1}U.
\end{equation}
For convenience we also take into account possible screening with
screening length \( \kappa  \), i.e.
\begin{equation}
U_{ij}\rightarrow U_{ij}e^{-\frac{\left| z_{i}-z_{j}\right| }{\kappa }}.
\end{equation}
Under the transformation Eq.~(\ref{invar_u}) the screening length should be 
multiplied by \( \Lambda \). In Fig.~\ref{lambde} the
parameter $\lambda$ is plotted for some typical cases, showing its dependence
on wire width.

\section{Two-electron wave functions}

Wave function for the case of two electrons
are expressed in terms of a set of operators 
\( c_{ij}^{\left( S,S_{z}\right) \dagger } \) creating an electron pair 
at sites  \( i \) and \( j \) with spin 
\( S \) and $z$-component \( S_{z} \), i.e.
\begin{equation}
\label{psi2el}
\left| \psi \right\rangle =\sum _{ij}\psi _{ij}c_{ij}^{\left(
S,S_{z}\right) 
\dagger }\left| 0\right\rangle .
\end{equation}
The base states
\begin{subequations}

\begin{eqnarray}
c_{ij}^{\left( 0,0\right) \dagger }\left| 0\right\rangle  & = & 
\frac{c_{i\uparrow }^{\dagger }c_{j\downarrow }^{\dagger
}-c_{i\downarrow }^{\dagger }
c_{j\uparrow }^{\dagger }}{\sqrt{2}}\left| 0\right\rangle ,\\
c_{ij}^{\left( 1,1\right) \dagger }\left| 0\right\rangle  & = 
& c_{i\uparrow }^{\dagger }c_{j\uparrow }^{\dagger }\left| 0\right\rangle ,\\
c_{ij}^{\left( 1,0\right) \dagger }\left| 0\right\rangle  & = & 
\frac{c_{i\uparrow }^{\dagger }c_{j\downarrow }^{\dagger }+
c_{i\downarrow }^{\dagger }c_{j\uparrow }^{\dagger }}{\sqrt{2}}\left| 0\right\rangle ,\\
c_{ij}^{\left( 1,-1\right) \dagger }\left| 0\right\rangle  & = 
& c_{i\downarrow }^{\dagger }c_{j\downarrow }^{\dagger }\left| 0\right\rangle .
\end{eqnarray}
\end{subequations}
form a complete set.

If $\left| \psi \right\rangle$ is a solution of  Schr\"odinger equation
\begin{equation}
H\left| \psi \right\rangle =E\left| \psi \right\rangle ,
\end{equation}
then the coefficients \( \widetilde{\psi} _{ij} \) solve the system of 
linear equations
\begin{eqnarray}
\label{sistem2el}
&&t\left( \widetilde{\psi }_{i-1j}+\widetilde{\psi
}_{i+1j}+\widetilde{\psi }_{ij-1}+
\widetilde{\psi }_{ij+1}\right) 
=\nonumber\\&&\quad\quad=\left( \epsilon _{i}+\epsilon
_{j}+U_{ij}-E\right) 
\widetilde{\psi }_{ij},
\end{eqnarray}
where we use compact notation
\begin{equation}
\widetilde{\psi }_{ij}=\frac{1}{\sqrt{2}}\left( \psi _{ij}+(-1)^{S}\psi _{ji}\right) .
\end{equation}
In the basis Eq.~(\ref{psi2el}) the number of electrons on site $i$ is
\begin{equation}
\label{density}
\left\langle \psi \left| n_{i}\right| \psi \right\rangle =2\sum
_{j}\left| 
\widetilde{\psi }_{ij}\right| ^{2},
\end{equation}
the current for sites  \( i \) and \( i+1 \) is
\begin{equation}
\left\langle \psi \left| j^{i,i+1}\right| \psi \right\rangle =
-\frac{4t\Delta }{\hbar }\textrm{Im}\sum _{j}\widetilde{\psi }_{ij}^{\ast }
\widetilde{\psi }_{i+1j},
\end{equation}
the energy is
\begin{eqnarray}
&&\left\langle \psi \left| H\right| \psi \right\rangle =\nonumber\\
&&\quad\quad-t\sum_{ij} \widetilde{\psi }_{ij}^{\ast }
\left( \widetilde{\psi }_{i+1j}+\widetilde{\psi }_
{i-1j}
+
\widetilde{\psi }_{ij+1}+\widetilde{\psi }_{ij-1}
\right)+\nonumber\\
&&\quad\quad+\sum _{ij}\left( \epsilon _{i}+\epsilon _{j}+
U_{ij}\right) \left| \widetilde{\psi }_{ij}\right| ^{2},
\end{eqnarray}
and the norm of the wave function Eq.~(\ref{psi2el}) is given with
\begin{equation}
\left\langle \psi |\psi \right\rangle =\sum _{ij}\left| 
\widetilde{\psi }_{ij}\right| ^{2}.
\end{equation}

We consider quantum wires which are almost perfect but for which there
is a very weak effective potential, giving rise to bound states.  
The cross-sections of these wires are sufficiently small that
the lowest transverse channel approximation is adequate for the energy
range of interest. The smooth variation in
cross-section also guarantees that inter-channel mixing is
negligible. We study here only wires with one weak bulge around \( z=0
\). There should exist single particle bound states of the system and  \(
E_{1}^{\left( \alpha \right) } \) is the energy of the state \( \alpha
\). The energy of two electron states is shifted and 
defined to be zero, if one electron is
bound and the other at the bottom of the single electron band, i.e.
\begin{equation}
\label{e2dogovor}
E\rightarrow E-E_{1}^{\left( 1\right) }.
\end{equation}
With this definition, the energy of two bound electrons is  negative
whereas it is positive when only one electron is bound.

\section{Hartree-Fock approximation}

\begin{figure}[htb]
{\par\centering \resizebox*{0.4\columnwidth}{!}{\includegraphics{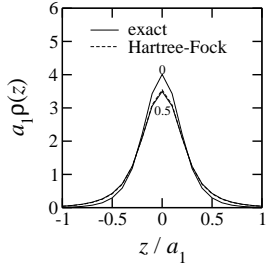}} \par}
\caption{\label{dens1} Electron density for singlet states with different \protect\(
\gamma \protect \). Full line corresponds to exact results and dashed line to 
the Hartree-Fock
approximation. Parameters as in Fig.~\ref{1_24}.}
\end{figure}

Here we neglect the Coulomb interaction between electrons
($\gamma=0$). In the ground state both electrons are in 
the same state \( \left| \varphi \right\rangle  \) and
the singlet wavefunction is
\begin{equation}
\left| \psi \right\rangle =\frac{1}{\sqrt{2}}\sum _{ij}
\varphi _{i}\varphi _{j}c_{ij}^{\left( 0,0\right) \dagger }\left| 0\right\rangle 
.\label{HFwv}
\end{equation}
For finite $\gamma$ the best one-electron wavefunctions, \( \varphi _{i} \), are  
determined by minimizing the energy,
\begin{equation}
\frac{\partial }{\partial \varphi ^{\ast }_{i}}\frac{\left\langle \psi 
\left| H\right| \psi \right\rangle }{\left\langle \psi |\psi
\right\rangle }=0.\label{HFmin}
\end{equation}
Which leads to the equation
\begin{equation}
\label{varprinc}
\frac{\partial }{\partial \varphi ^{\ast }_{i}}\left\langle 
\psi \left| H\right| \psi \right\rangle -\left( E+E_{1}^{\left( 
1\right) }\right) \frac{\partial }{\partial \varphi ^{\ast }_{i}}
\left\langle \psi |\psi \right\rangle =0,
\end{equation}
where we have set
\begin{equation}
\frac{\left\langle \psi 
\left| H\right| \psi \right\rangle }{\left\langle \psi |\psi
\right\rangle }=E+E^{(1)}_1,\label{php}
\end{equation}
taking into account the energy shift Eq.~(\ref{e2dogovor}).
The expectation value of energy and the norm is 
\begin{equation}
\left\langle \psi \left| H\right| \psi \right\rangle =
2\left\langle \varphi \left| H_{1}\right| \varphi 
\right\rangle \left\langle \varphi |\varphi \right
\rangle +\sum _{ij}U_{ij}\left| \varphi _{i}\right| ^{2}\left| \varphi _{j}\right| ^{2},
\end{equation}
and
\begin{equation}
\left\langle \psi |\psi \right\rangle =\left\langle \varphi |\varphi \right\rangle ^{2}.
\end{equation}
From Eq.~(\ref{varprinc}) follows a system of equations for coefficients \( \varphi _{i} \)
\begin{equation}
\left\langle 0\left| c_{i}H_{1}\right| \varphi \right\rangle +\sum
_{j}U_{ij}\left| 
\varphi _{j}\right| ^{2}\varphi _{i}=E_{\textrm{hf}}\varphi _{i},
\end{equation}
where
$\left\langle 0\left| c_{i}H_{1}\right| \varphi \right\rangle=
-t(\varphi_{i-1}+\varphi_{i+i})+\epsilon_i \varphi$ is a one-electron tight-binding
Hamiltonian, $\sum _{j}U_{ij}\left| 
\varphi _{j}\right| ^{2}$ is a Hartree potential and 
\begin{equation}
E_{\textrm{hf}}=E+E_{1}^{\left( 1\right) }-\frac{\left\langle \varphi 
\left| H_{1}\right| \varphi \right\rangle }{\left\langle \varphi |\varphi \right\rangle }
\end{equation}
is the so-called Hartree-Fock energy.
The energy of a bound state is then given by
\begin{equation}
E=2E_{\textrm{hf}}-\sum _{ij}U_{ij}\left| \varphi _{i}\right|
^{2}\left| 
\varphi _{j}\right| ^{2}-E_{1}^{\left( 1\right) },
\end{equation}
where, due to double-counting of the interactions in single electron energies,
\( E_{\textrm{hf}} \) is subtracted.
In Fig.~\ref{dens1} electron density for singlet states with different \protect\(
\gamma \protect \) is presented for a shorter bulge with parameters as in Fig.~\ref{1_24}.

In the case of triplet two-electron states, 
the single electron states are different, \( \left|
\varphi \right\rangle  \) and \( \left| \bar{\varphi }\right\rangle
\). Choosing these to be orthogonal, we get

\begin{equation}
\left| \psi \right\rangle =\sum _{ij}\varphi _{i}
\bar{\varphi }_{j}c_{ij}^{\left( 1,1 \right) \dagger }\left| 0\right\rangle .
\end{equation}
The energy is now

\begin{eqnarray}
\left\langle \psi \left| H\right| \psi \right\rangle &=&
\left\langle \varphi \left| H_{1}\right| \varphi 
\right\rangle \left\langle \bar{\varphi }|\bar{\varphi }\right\rangle
\\ \nonumber
&+&
\left\langle \bar{\varphi }\left| H_{1}\right| \bar{\varphi
}\right\rangle 
\left\langle \varphi |\varphi \right\rangle 
+\frac{1}{2}\sum
_{ij}U_{ij}
\left| \varphi _{i}\bar{\varphi }_{j}-\bar{\varphi }_{i}\varphi _{j}\right| ^{2},
\end{eqnarray}
and the norm is
\begin{equation}
\left\langle \psi |\psi \right\rangle =\left\langle \varphi 
|\varphi \right\rangle \left\langle \bar{\varphi }|\bar{\varphi }\right\rangle ,
\end{equation}
The system of equations for the coefficients \( \varphi _{i} \) (and
equivalent for \( \bar{\varphi }_{i} \)) is
\begin{equation}
\left\langle 0\left| c_{i}H_{1}\right| \varphi _{i}\right\rangle +
\sum _{j}U_{ij}\left| \bar{\varphi }_{j}\right| ^{2}\varphi _{i}-
\sum _{j}U_{ij}\bar{\varphi }_{j}^{\ast }\bar{\varphi }_{i}\varphi _{j}=
E_{\textrm{hf}}\varphi _{i},\label{d13}
\end{equation}
where
\begin{equation}
E_{\textrm{hf}}=E+E_{1}^{\left( 1\right) }-\frac{\left\langle 
\bar{\varphi }\left| H_{1}\right| \bar{\varphi }\right\rangle }
{\left\langle \bar{\varphi }|\bar{\varphi }\right\rangle },
\end{equation}
Eq.~(\ref{d13})
is a single-particle tight-binding Sch\" odinger equation with Hamiltonian

\begin{equation}
\tilde{H}_{1}=-\sum _{i}\tilde{t}_{ij}\left( 
c_{i}^{\dagger }c_{j}+c_{j}^{\dagger }c_{i}\right) +\sum _{i}\tilde{\epsilon }_
{i}c_{i}^{\dagger }c_{i}
\end{equation}
with  potential
\begin{equation}
\tilde{\epsilon }_{i}=\epsilon _{i}+\sum _{j}U_{ij}\left| \bar{\varphi }_{j}\right| ^{2}
\end{equation}
and renormalized hoping parameters
\begin{equation}
\tilde{t}_{ij}=t\delta _{j,i\pm 1}+\sum _{j}U_{ij}\bar{\varphi }_{j}^{\ast }
\bar{\varphi }_{i}.
\end{equation}
The energy of the triplet bound state is then
\begin{equation}
E=E_{\textrm{hf}}+\bar{E}_{\textrm{hf}}-\frac{1}{2}\sum
_{ij}U_{ij}\left| 
\varphi _{i}\bar{\varphi }_{j}-\bar{\varphi }_{i}\varphi _{j}\right| ^{2}-E_{1}^{
\left( 1\right) }.
\end{equation}

\begin{figure}[htb]
{\par\centering \resizebox*{0.9\columnwidth}{!}{\rotatebox{0}
{\includegraphics{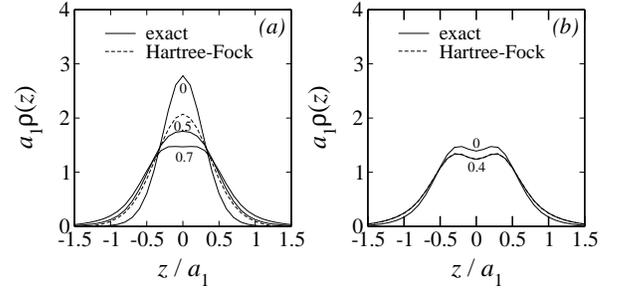}}} \par}
\caption{\label{dens2}
Electron singlet (a) and triplet (b) state density for various \protect\( \gamma
\protect \). Parameters are as in Fig.~\ref{1_06}. }
\end{figure}

In Fig.~\ref{dens2} the electron density for singlet (a) and 
for triplet (b) states are shown for different
\protect\(\gamma \protect \). Other parameters are taken as in the case 
of the longer bulge, Fig.~\ref{1_06}.  As discussed in the text relating to
Fig.~\ref{1_06}, Hartree-Fock approximation for the singlet is less reliable since the
Coulomb repulsion is stronger due to both electrons being in the same state. Indeed,
for  $\gamma ~ \sim 0.7$ the Hartree-Fock approximation does not yield the bound state 
found in the exact result.

\acknowledgments
The authors wish to acknowledge N.J. Appleyard,  A.V. Khaetskii,  C.J. 
Lambert, M. Pepper and K.J. Thomas for helpful discussions. This work was 
supported by the EU and the MoD.


\begin{references} 

\bibitem{walther92} {M. Walther, E. Kapon, D.M. Hwang, E. Colas, and
L. Nunes, Phys. Rev. B {\bf 45}, 6333 (1992); M. Grundmann {\em et
al.}, Semicond. Sci. Tech. {\bf 9}, 1939 (1994); R. Rinaldi {\em et
al.}, Phys. Rev. Lett. {\bf 73}, 2899 (1994).}

\bibitem{ramvall97} {P. Ramvall {\em et al.}, Appl. Phys. Lett. {\bf
71}, 918 (1997).}

\bibitem{yacoby96} {A. Yacoby {\em et al.}, Phys. Rev. Lett. {\bf 77},
4612 (1996).}

\bibitem{kristensen98} {A. Kristensen {\em et al.}, Contributed paper
for ICPS24, Jerusalem, August 2-7, 1998.}

\bibitem{wees88} {B.J. van Wees {\em et al.}, Phys. Rev. Lett. {\bf
60}, 848 (1998).}

\bibitem{wharam88} {D.A. Wharam {\em et al.}, J. Phys. C: Solid state
Phys. {\bf 21}, L209 (1988).}

\bibitem{houten92} {H. van Houten, C.W.J. Beenakker, and B.J. van
Wees, in {\it Semiconductors and Semimetals}, Vol. {\bf 35}, edited by
M.A. Reed (Academic Press, New York, 1992).}

\bibitem{thomas96} {K.J. Thomas {\em et al.}, Phys. Rev. Lett. {\bf
77}, 135 (1996); Phys. Rev. B {\bf 58}, 4846 (1998); Phys. Rev. B {\bf
59}, 12252 (1999).}

\bibitem{liang} {C.-T. Liang {\em et al.}, Phys. Rev. B {\bf 60}, 10687
(1999).}

\bibitem{pyshkin} {K.S. Pyshkin {\em et al.}, Phys. Rev. B {\bf 62}, 15842
(2000).}

\bibitem{kaufman99} {D. Kaufman {\em et al.}, Phys. Rev. B {\bf 59},
R10433 (1999).}

\bibitem{nuttinck} {S. Nuttinck {\em et al.}, Jpn. J. Appl. Phys. {\bf 39},
655 (2000); K. Hashimoto {\em at al.}, Jpn. J. Appl. Phys. {\bf 40},
3000 (2001).}

\bibitem{kristensen00} {A. Kristensen {\em et al.}, Phys. Rev. B {\bf 62},
10950 (2000).}

\bibitem{liang00} {C.-T. Liang {\em et al.},  Phys. Rev. B {\bf 61}, 9952
(2000).}

\bibitem{kondo2002} {C.M. Cronenwett {\em et al.},
Phys. Rev. Lett. {\bf 88}, 226805 (2002)}

\bibitem{appleyard00} {N.J. Appleyard {\em et al.}, Phys. Rev. B {\bf
62}, R16275 (2000).}

\bibitem{maslov95} {D.L. Maslov, Phys. Rev. B {\bf 52}, R14368,
1995}.

\bibitem{wang98} {Chuan-Kui Wang and K.-F. Berggren, Phys. Rev. B {\bf
57}, 4552 (1998).}

\bibitem{fasol94} {G. Fasol and H. Sakaki, Jpn. J. Appl.  Phys. {\bf
33}, 879 (1994).}

\bibitem{rejec002d} {T. Rejec, A. Ram\v sak, and J.H. Jefferson,
J. Phys., Condens. Matter {\bf 12}, L233 (2000).}

\bibitem{rejec003d} {T. Rejec, A. Ram\v sak, and J.H. Jefferson,
Phys. Rev. B {\bf 62}, 12985 (2000).}

\bibitem{rejec02} {T. Rejec, A. Ram\v sak, and J.H. Jefferson,
Phys. Rev. B {\bf 65}, 235301 (2002).}

\bibitem{flambaum00} {V.V. Flambaum and M.Yu. Kuchiev, Phys. Rev. B
{\bf 61}, R7869 (2000).}

\bibitem{landau} {H. Bruus, V. Cheianov, and K. Flensberg, Physica E
{\bf 10}, 97 (2001).}

\bibitem{sushkov1} {O.P. Sushkov, Phys. Rev. B {\bf 64}, 155319
(2001)}

\bibitem{tokura} {Y. Tokura and A. Khaetskii, Physica E {\bf 12}, 711 (2002).}

\bibitem{meir2002} {Y. Meir, K. Hirose, and N.S. Wingreen, \\cond-mat/0207044.}

\bibitem{meirav91} {U. Meirav {\em et al.}, Z. Phys. B {\bf 85}, 357
(1991).}

\bibitem{rejecand} {T. Rejec, A. Ram\v sak, and J.H. Jefferson, in
{\it Kondo Effect and Dephasing in Low-Dimensional Metallic Systems},
edited by V. Chandrasekhar, C. Van Haesendonck, and A. Zawadowski,
NATO ARW, Ser. II, Vol. {\bf 50} (Kluwer, Dordrecht, 2001).}

\bibitem{anderson61} {P.W. Anderson, Phys. Rev. {\bf 124}, 41 (1961); 
G.D. Mahan, {\it Many-Particle Physics}, Plenum
Press, New York (1990).} 

\bibitem{jauregui} {K. Jauregui, W. H\" ausler, and B. Kramer, Euriphys.
Lett. {\bf 24} 581 (1993); D.L.J. Tipton, PhD Thesis, King's College London (2001).} 

\bibitem{landauQM} {L.D. Landau and E.M. Lifshitz, {\it Quantum
Mechanics} (Pergamon Press, Oxford, 1977).}

\bibitem{oppenheimer28} {J.R. Oppenheimer, Phys. Rev. {\bf 32}, 361
(1928); N.F. Mott, Proc. Roy. Soc. A {\bf 126}, 259 (1930).}

\bibitem{abramovitz} {M. Abramowitz and I.A. Stegun, Handbook of
mathematical functions, (Dover publications, Inc., New York).}

\bibitem{landauer57} {R. Landauer, IBM J. Res. Dev. {\bf 1}, 223
(1957); {\bf 32}, 306 (1988); M. B\"{u}ttiker, Phys. Rev. Lett. {\bf
57}, 1761 (1986).}

\bibitem{ramsak98} {A. Ram\v sak, T. Rejec, and J.H. Jefferson,
Phys. Rev. B {\bf 58}, 4014 (1998).}

\bibitem{patel91} {N.K. Patel {\em et al.}, Phys. Rev. B {\bf 44},
13549 (1991); K.J. Thomas {\em et al.}, Phil. Mag. B {\bf 77}, 1213
(1998).}

\bibitem{lieb62} {E. Lieb and D. Mattis, Phys. Rev. {\bf 125}, 164
(1962).}

\bibitem{anddot} {L.I. Glazman and M.E. Raikh, JETP Lett. {\bf 47},
452 (1988); T.K. Ng and P.A. Lee, Phys. Rev. Lett {\bf 61}, 1768 (1988).}

\bibitem{jeff} {C.E. Creffield, J.H. Jefferson, Sarben Sarkar, and
D.L.J. Tipton, Phys. Rev. B {\bf 62}, 7249 (2000).}

\bibitem{hewsonbook} {see, e.g., A.C. Hewson, {\it The Kondo problem
to heavy fermions}, (Cambridge university press, Cambridge, 1997).}

\bibitem{rejecram} {T. Rejec and A. Ram\v sak, unpublished.}

\bibitem{boese01} {D. Boese and R. Fazio, Europhys. Lett. {\bf 56},
576 (2001).}

\end{references}
\end{document}